\documentclass[tighten, notrackchanges, twocolumn]{aastex631}
\usepackage{amsmath}
\usepackage{amssymb}
\usepackage{amssymb}
\usepackage{mathtools}
\usepackage{mathrsfs}
\usepackage{enumerate}
\usepackage{morefloats}
\usepackage{longtable}
\usepackage{afterpage}
\usepackage{acronym}   
\hypersetup{linkcolor=red,citecolor=blue,urlcolor=magenta}
\usepackage{enumitem}
\usepackage{soul}

\usepackage[makeroom]{cancel}

\graphicspath{{figs/}}

\usepackage{xspace}

\newcommand*{\XPSI}{X-PSI\xspace}
\newcommand*{\NICER}{{NICER}\xspace}
\newcommand*{\xmm}{{XMM-Newton}\xspace}
\newcommand*{\MultiNest}{\textsc{MultiNest}\xspace}

\newcommand{\joo}{PSR~J0030$+$0451\xspace}
\newcommand{\joh}{PSR~J0740$+$6620\xspace}
\newcommand{\joA}{PSR~J0437$-$4715\xspace}

\acrodef{X-PSI}{X-ray Pulse Simulation and Inference}
\acrodef{MSP}{millisecond pulsar}
\acrodef{PPM}{Pulse Profile Modeling}
\acrodef{EoS}{equation of state}
\acrodef{ISS}{International Space Station}
\acrodef{NS}{neutron star}
\acrodefplural{NS}[NSs]{neutron stars}
\acrodef{NICER}{Neutron Star Interior Composition Explorer}
\acrodef{GTI}{Good Time Interval}
\acrodefplural{GTI}[GTIs]{Good Time Intervals}

\newcommand{\be}{\begin{equation}}
\newcommand{\ee}{\end{equation}}

\usepackage{enumitem,amssymb}
\usepackage{tikz}
\def\checkmark{\tikz\fill[scale=0.4](0,.35) -- (.25,0) -- (1,.7) -- (.25,.15) -- cycle;} 

\defcitealias{Riley2019}{R19}
\defcitealias{Vinciguerra2023a}{V23a}

\received{}
\revised{}
\accepted{}
\published{}

\submitjournal{nowhere}

\shorttitle{NICER updates for PSR J0030+0451}
\shortauthors{Vinciguerra~et~al.}

\begin{document}

\title{An updated mass-radius analysis of the 2017-2018 NICER data set of PSR J0030+0451 \footnote{Submitted on -, 7, 2023}}

\correspondingauthor{S.~Vinciguerra}
\email{s.vinciguerra@uva.nl}

\author[0000-0003-3068-6974]{Serena~Vinciguerra}
\affil{Anton Pannekoek Institute for Astronomy, University of Amsterdam, Science Park 904, 1098XH Amsterdam, the Netherlands}

\author[0000-0001-6356-125X]{Tuomo~Salmi}
\affil{Anton Pannekoek Institute for Astronomy, University of Amsterdam, Science Park 904, 1098XH Amsterdam, the Netherlands}

\author[0000-0002-1009-2354]{Anna~L.~Watts}
\affil{Anton Pannekoek Institute for Astronomy, University of Amsterdam, Science Park 904, 1098XH  Amsterdam, the Netherlands}

\author[0000-0002-2651-5286]{Devarshi~Choudhury}
\affil{Anton Pannekoek Institute for Astronomy, University of Amsterdam, Science Park 904, 1098XH Amsterdam, the Netherlands}

\author[0000-0001-9313-0493]{Thomas~E.~Riley}
\affil{Anton Pannekoek Institute for Astronomy, University of Amsterdam, Science Park 904, 1098XH Amsterdam, the Netherlands}

\author[0000-0002-5297-5278]{Paul~S.~Ray}
\affil{Space Science Division, U.S. Naval Research Laboratory, Washington, DC 20375, USA}

\author[0000-0002-9870-2742]{Slavko~Bogdanov} 
\affil{Columbia Astrophysics Laboratory, Columbia University, 550 West 120th Street, New York, NY 10027, USA}

\author[0000-0002-0428-8430]{Yves~Kini}
\affil{Anton Pannekoek Institute for Astronomy, University of Amsterdam, Science Park 904, 1098XH Amsterdam, the Netherlands}

\author[0000-0002-6449-106X]{Sebastien~Guillot}
\affil{Institut de Recherche en Astrophysique et Plan\'{e}tologie, UPS-OMP, CNRS, CNES, 9 avenue du Colonel Roche, BP 44346, F-31028 Toulouse Cedex 4, France}

\author[0000-0001-8804-8946]{Deepto~Chakrabarty} 
\affil{Massachusetts Institute of Technology, Cambridge, MA, USA}

\author[0000-0002-6089-6836]{Wynn~C.~G.~Ho}
\affil{Department of Physics and Astronomy, Haverford College, 370 Lancaster Avenue, Haverford, PA 19041, USA}

\author[0000-0002-1169-7486]{Daniela~Huppenkothen}
\affil{SRON Netherlands Institute for Space Research, Niels Bohrweg 4, NL-2333 CA Leiden, the Netherlands}

\author[0000-0003-4357-0575]{Sharon~M.~Morsink}
\affil{Department of Physics, University of Alberta, 4-183 CCIS, Edmonton, AB, T6G 2E1, Canada}

\author[0000-0002-9249-0515]{Zorawar~Wadiasingh}
\affiliation{Department of Astronomy, University of Maryland, College Park, Maryland 20742, USA}
\affiliation{Astrophysics Science Division, NASA Goddard Space Flight Center, Greenbelt, MD 20771, USA.}
\affiliation{Center for Research and Exploration in Space Science and Technology, NASA/GSFC, Greenbelt, Maryland 20771, USA}

\author[0000-0002-4013-5650]{ Michael~T.~Wolff}
\affil{Space Science Division, U.S. Naval Research Laboratory, Washington, DC 20375, USA}

\begin{abstract}
In 2019 the \NICER collaboration published the first mass and radius inferred for \joo, thanks to \NICER observations, and consequent constraints on the equation of state characterising dense matter. Two independent analyses found  a mass of $\sim 1.3-1.4\,\mathrm{M_\odot}$ and a radius of $\sim 13\,$km. They also both found that the 
hot spots were all located on the same hemisphere, opposite to the observer, and that at least one of them had a significantly elongated shape. Here we reanalyse, in greater detail, the same \NICER data set, incorporating the effects of an updated \NICER response matrix and using an upgraded analysis framework. We expand the adopted models and jointly analyse also \xmm data, which enables us to better constrain the fraction of observed counts coming from \joo. Adopting the same models used in previous publications, we find consistent results, although with more stringent inference requirements. We also find a multi-modal structure in the posterior surface. This becomes crucial when \xmm data is accounted for. 
Including the corresponding constraints disfavors the main solutions found previously, in favor of the new and more complex models.  These have inferred masses and radii of $\sim [1.4 \mathrm{M_\odot}, 11.5$ km] and $\sim [1.7 \mathrm{M_\odot}, 14.5$ km], depending on the assumed model. They display configurations that do not require the two hot spots generating the observed X-rays to be on the same hemisphere, nor to show very elongated features, and point instead to the presence of temperature gradients and the need to account for them.

\end{abstract}
\keywords{}

\section{Introduction} \label{sec:intro}
The Neutron Star Interior Composition Explorer (\NICER) is an instrument installed on the International Space Station to detect the soft thermal X-ray emission of rotation-powered \acp{MSP}. By modeling this emission, originating from the hot polar caps, we are able to infer \ac{NS} masses and radii, hence constraining the \ac{EoS} governing dense and cold matter. 
So far the \NICER collaboration has released results for two sources: \joo \citep[][]{Miller2019,Riley2019} and the high mass pulsar \joh \citep[][]{Miller2021,Riley2021,Salmi2022}. 

The X-ray emission of the \acp{MSP} that \NICER targets is characterised by pulsations generated by return currents heating up the \ac{NS} surface at the magnetic poles. Special and general relativistic effects on the propagation of radiation ensure that the phase-resolved spectrum of the pulsations carries information about the space-time surrounding the \ac{NS}. 
By modeling all of the relevant relativistic effects, a technique known as \ac{PPM}, 
it is possible to infer mass and radius \citep[for a general introduction to \ac{PPM} see][]{Watts2019,Bogdanov2019b,Bogdanov2021}. As a by-product of the analysis,  \ac{PPM} also allows us to infer the properties (size, shape, location) of the hot emitting regions.

In this paper we revisit the first source for which \NICER results were released, \joo, using the \ac{X-PSI}\footnote{\url{https://github.com/xpsi-group/xpsi}} package \citep{xpsi}.   Our analysis is an update to \citet[][hereafter R19]{Riley2019}, since validated by \citet{Afle23} who also used \XPSI; the independent analysis of \citet{Miller2019} used a different pipeline.  

\joo is an isolated \ac{MSP} rotating at $\sim 205$\, Hz, located at a distance of $\sim 325$ pc \citep{Arzoumanian2018,Ding23}. 
The first \XPSI analysis of the \NICER data set for \joo \citepalias[][]{Riley2019} found that several models for the size and shape of the hot polar caps could reproduce the data quite well. However, when the models were ranked according to their evidence, the preferred model was one in which emission originated from one small (circular) and one extended (arc-like) hot-spot,  located in the same rotational hemisphere. 
An independent analysis of the same data by \citet{Miller2019}, using a different parameter estimation code, found qualitatively similar spot shapes and locations on the star, although they used ovals instead of arcs. 
The shape and location of the hot emitting regions suggested 
the need for profound changes to the standard picture of the \acp{NS}'s magnetic field, as they were incompatible with a classical centered dipole \citep{Bilous_2019,Chen2020,Kalapotharakos2021}.

For the preferred emission pattern and viewing geometry, the mass and radius inferred by \citetalias{Riley2019} for \joo was $1.34^{+0.15}_{-0.16}\,\mathrm{M_\odot}$ and $12.71^{+1.14}_{-1.19}$\, km, where the uncertainties, here and throughout this work, are specified at approximately the 16\% and 84\% quantiles in the 1D marginal posterior (for comparison, \citealt{Miller2019} inferred $1.44^{+0.15}_{-0.14}\,\mathrm{M_\odot}$ and $13.02^{+1.24}_{-1.06}$\, km). The mass and radius posteriors have since been used in a variety of studies to constrain the \ac{EoS} \citep[see for example][]{Raaijmakers2021,TangSP21,Biswas2022,Sabatucci22,Rutherford23,SunX23}. The fact that \joo has an inferred radius similar to that inferred for the $\sim 2.1\,\mathrm{M_\odot}$ pulsar \joh \citep{Fonseca20,Miller2021,Riley2021,Salmi2022}, despite the much lower inferred mass, is particularly notable \citep{Raaijmakers2021}. 
The shift in radius as mass increases can be used to distinguish between \ac{EoS} models \citep{Drischler21}.

Improving the robustness of the \NICER \ac{PPM} results and updating them whenever possible is therefore crucial for advancing our understanding of the \ac{EoS}. In a companion paper \citep[][hereafter V23a]{Vinciguerra2023a}  
parameter recovery simulations, tailored to \joo, were carried out to test the reliability of our results for complex hot spot models, and the sensitivity to stochastic fluctuations and inference setups. 
This study revealed some sensitivity of \ac{PPM} posteriors to noise, analysis settings and random sampling processes. Their effects are interconnected: depending on the noise realisation, specific analysis settings may or may not be sufficient to assert a certain degree of stability in the results and, in general, computationally cheap analysis settings are more prone to variability due to random sampling processes. 
The other major finding of this work was the presence of a clearly multi-modal structure in the posterior. 
These posterior maxima are often considerably different in terms of associated probabilities, so they do not always appear as clear multi-modal structures in the posterior plots. Sometimes they may be present as tails of the distributions, other times they may even be completely obliterated by the main mode. However their properties can be considerably different, and the application of independent constraints may completely shift their importance.  

Due to limitations in the computational resources, the \citetalias{Vinciguerra2023a} simulation study focused on synthetic data sets based on only two different parameter vectors. 
This implies that some of its findings may not be general.
In particular, comparing the results of \citetalias{Vinciguerra2023a} with the ones reported in \citetalias{Riley2019}, we 
note some
differences that may be tied to the specific choice of parameter vectors. 
For example, a negligible difference in evidence between different models was found by \citetalias{Vinciguerra2023a} but not by \citetalias{Riley2019}. 
\citetalias{Vinciguerra2023a} also reported that different models could recover consistent mass and radius, given a specific synthetic data set; while \citetalias{Riley2019} showed that when analysing the original \NICER data set, presented in \citet{Bogdanov19a}, employing different models drastically changes the inferred mass and radius posterior distributions.  In this paper we apply 
what we have learnt from \citetalias{Vinciguerra2023a}
and revisit the \ac{PPM} for \joo, making a number of improvements over the analysis presented in \citetalias{Riley2019}. 

\XPSI has undergone a series of major updates since the analysis presented in \citetalias{Riley2019}: for example, the surface pattern model suite is more comprehensive,
and it is now possible to perform joint analysis with data sets observed by multiple instruments.  
In \citetalias{Riley2019}, the inferred background was checked a posteriori against constraints derived from \xmm observations \citep{BogdanovGrindlay2009}.  In this paper we include 
\xmm constraints directly in the inference analysis, as was done for \joh in \citet{Riley2021,Miller2021,Salmi2022}. With more computational resources, we were also able to carry out a wider range and higher resolution study.  This enables us to explore the robustness of the findings in more depth. 
Another improvement is to the instrument response model of \NICER, which has been updated since the analysis presented in \citetalias{Riley2019}. What we present here is the analysis of a revised \NICER data set (hereafter B19v006), derived from the one presented in \citet{Bogdanov19a} and covering the same observations, but modified to be consistent with the latest \NICER response matrix. This therefore becomes the baseline to which inference of \joo data sets containing newer observations should eventually be compared. 

Below, for our \ac{PPM} inferences, we will consider four different models which describe \joo's system, adopting increasingly more complex surface patterns. 
With \NICER-only analyses, we test the sensitivity of the identified posteriors to various assumptions and analysis settings.
We also check the impact of random sampling processes, by repeating multiple times inferences identically set up. 
Given the various analyses initiated with \NICER-only data, we decided to appoint four reference runs, one for each of the adopted models. 
These are used for the general discussion and have been performed with computationally expensive analysis settings; however, the robustness of our findings is also in this case not always guaranteed. 
We only performed one production run per model, when jointly analysing \NICER and \xmm data; in this case there is therefore no need for references. 
Since joint inferences require considerably higher computing resources, the analysis settings are, most of the time, less optimal compared to the \NICER-only case. 
The corresponding posteriors are therefore not proven to be robust, yet. 
To help and guide the interpretation of our findings, 
more details and a wider context are given throughout the text that follows. 

In Section \ref{sec:method} we describe the models adopted for this analysis and the most significant upgrades to \XPSI since the \citetalias{Riley2019} analysis.  In Section \ref{sec:dataset} we outline the most relevant properties and changes to the  \NICER and \xmm data sets that are analyzed. The results of our analysis are then presented in Section \ref{sec:results} and discussed in Section \ref{sec:discussion}. We present our final remarks in Section \ref{sec:conclusion}.

\section{Methodology: \XPSI upgrades and main setups} 
\label{sec:method}
In this work, we use the package \XPSI to analyse the X-ray emission produced by \joo and detected by \NICER. 
\NICER data are registered as events (counts) per PI (instrumental-energy) channel, where each event is characterised by a specific detection time. As described in \citetalias{Riley2019}, data collected over many ($\mathcal{O}(10^8)$) rotational cycles are folded over the spin period of the \ac{MSP} of interest, in our case \joo, and are then binned into 32 phase bins. 
The data analyzed by \XPSI therefore have  the form of number of events (counts) per instrumental channel and rotational phase bin.

\XPSI is a software package which performs parameter estimation by modeling the thermal emission generated at the \ac{NS} surface and detected by \NICER, 
following the methodology described in \citet[][]{Bogdanov2019b,Riley2019,Bogdanov2021, Riley2021}. 
Each hot spot on the \ac{NS} surface is modeled by overlapping spherical caps (see Section \ref{subsec:analysis} for more details). 
To account for the potential well and the shape of the \ac{NS}, our analysis relies on relativistic ray-tracing described by the Oblate Schwarzschild plus Doppler approximation, introduced by \citet[][]{Morsink2007,AGM14}. 
The final intensity at the observer is then calculated accounting for the effects of the \ac{NS} atmosphere and the interstellar medium.  Such a signal is generated for every parameter vector sampled by \MultiNest \citep{Feroz2008,Feroz2009,Feroz2019}, the algorithm adopted within \XPSI to explore the model parameter space through nested sampling algorithm \citep{Skilling2004}. 
This simulation process is essential for our inference analysis, which then compares these synthetic signals against the actual \NICER data within a Bayesian framework.

In this work we use \XPSI versions \texttt{v0.7} and \texttt{v1} (for inference \texttt{v0.7.3}, \texttt{v0.7.9} and \texttt{v0.7.10} - which differ from one another only by upgrades and minor bug fixes that are not expected to affect the results - and \texttt{v1.0.0 - v2.0.0} for post-processing). 
Compared to the version used in \citetalias{Riley2019} (\XPSI \texttt{v0.1}) these \XPSI versions incorporate the possibility of multiple images, as described in Section 2.2.3 of \citet{Riley2021}. 
Note that we do not expect this modification to significantly affect our results, since it is relevant for compactness values $C = M R^{-1}> 0.284$ (for brevity, we assume $c=G=1$) and \citetalias[][]{Riley2019} found negligible posterior probability for compactness above $\sim 0.2$ in all tested models (see Figure 19 of that paper). 

\subsection{\XPSI Models}
\label{subsec:analysis}
A summary of the method underlying the analyses performed with \XPSI and its recent upgrades and modifications can be found in Section 2 of \citetalias{Vinciguerra2023a}. 
We expand on that overview by presenting below additional analysis specifications that are necessary to introduce the wider range of tests included in this work. 

\subsubsection{Atmosphere and Interstellar Medium}
\label{subsubsec:atmISM}
Throughout this work we assume that the hot spots of \joo have 
a fully ionized and non-magnetic \texttt{NSX} hydrogen atmosphere \citep{Ho01, HH09}. 
As in \citet{Riley2021}, \citet{Salmi2023} and \citetalias{Vinciguerra2023a},
the intensity of the radiation field is calculated by interpolating it from a table (extended compared to that used in \citetalias{Riley2019}), which expresses it for different values of effective temperature, surface gravity, photon energy and the cosine of emission angle calculated from the surface normal \citepalias[for more details see Section 2.4.1  of][]{Riley2019}.
A detailed discussion of the implications of  this choice (and possible alternatives) is presented in \citet[][]{Salmi2023}, specifically for \joo and the data set analyzed here. 
For the first time, we test the potential impact of adding possible in-band emission from the \ac{NS} surface, outside the hot spots.  For this preliminary test, we still assume a fully ionized hydrogen atmosphere for the hot spots, but model the radiation generated from the remaining part of the surface as black-body emission, to reduce the corresponding computational cost. Even with this simplification, adding such a component can indeed considerably increase the computational resources required \footnote{Setting what we consider an adequate number of cells to present such surface (\texttt{sqrt\_num\_cells = 512}), the core hours required for an \texttt{ST-U} inference run roughly quadrupled. For the \texttt{ST+PST} inference, we had to lower by a lot the number of cells (\texttt{sqrt\_num\_cells = 64}) and still also set the sampling efficiency to 0.8, to make the run affordable (about twice the core hours required with a similar run, excluding the emission from the remaining part of the \ac{NS} surface).}.

The attenuating effects that the interstellar medium has at different energies are simulated with the same pre-computed tables, based on the \texttt{tbnew} model \footnote{\url{https://pulsar.sternwarte.uni-erlangen.de/wilms/research/tbabs/}}, used in \citetalias{Riley2019}. 
As outlined in  \citetalias{Riley2019}, we parameterize the effect of the interstellar medium with a single variable, the neutral hydrogen column density $N_{\mathrm{H}}[\mathrm{cm}^{-2}]$. Other chemical abundances are then assumed from it, following \citet{Wilms2000}. 

\subsubsection{Parametrizations of Hot Spots}
\label{subsubsec:HSmodels}
To model the thermal emission of the hot spots in the \NICER band, 
we use parameterized models that are motivated by theoretical
studies of return currents and polar cap heating in rotation-powered pulsars \citep{Harding01,Harding11,Timokhin13,Kalapotharakos2014,Gralla17,Lockhart19,Kalapotharakos2021}. 
In particular we define each hot spot with one or two overlapping spherical caps on the \ac{NS} surface. 
One of these components is always emitting at a constant and uniform temperature; the other can also radiate (forming a hot spot with two temperatures) or mask the X-rays of the emitting component (widening the range of possible hot spot shapes to include arcs or rings). 

\XPSI analyses of \NICER data have always assumed the presence of two non-overlapping hot spots, motivated by their theoretical connection to the magnetic poles and the data structure of the pulse profile. 
We refer to these two hot spots as the {\it primary} and the {\it secondary}. 
Within the \XPSI framework we can define different models, given the set up just described. 
More details about the naming convention adopted for our models and a schematic representation of them is reported in Section 2.3.3 and Figure 1 of \citetalias{Vinciguerra2023a}. \\
In this work we adopt nested models that describe the surface patterns with increasing complexity:
\begin{itemize}[noitemsep]
    \item \texttt{\bf{ST-U}}: where each of the two hot spots is described by a single spherical cap. The primary is defined as the hot spot with lower colatitude with respect to the angular momentum vector (spin axis)\footnote{As in \citetalias{Riley2019}, the rotation is defined by the right-hand rule, with the thumb pointing at North.};
    \item  \texttt{\bf{ST+PST}}: where the primary (\texttt{ST}) is described by a single spherical cap and the secondary (\texttt{PST}) by two components, one emitting and one masking (see Figure 2 of \citetalias{Vinciguerra2023a}). Within this model, for comparison we adopt both the prior used for analyses in \citetalias{Riley2019} and the more Comprehensive Hot spot (CoH) prior described in Section 2.3.4 of \citetalias{Vinciguerra2023a}. The latter allows for the hot spot described by a single spherical cap to overlap with the masking component of the \texttt{PST} one; 
    \item {\texttt{\bf{ST+PDT}}: where the primary (\texttt{ST}) is described by a single spherical cap and the secondary (\texttt{PDT}) by two components, both emitting; 
    }
    \item \texttt{\bf{PDT-U}}: where each of the two hot spots are described by two emitting spherical caps. The primary is defined as the hot spot with lower colatitude.
\end{itemize}
The first two are described in detail in \citetalias{Vinciguerra2023a}; a more detailed explanation of the others is given below. 
\citetalias{Riley2019} suggested that one of the hot spots was already well represented by our simplest description (i.e. by a circular component with uniform temperature). 
For this reason, starting from the simplest model here considered, \texttt{ST-U}, we first increase the complexity of the description of only one of the hot spots: first allowing the creation of different shapes with the \texttt{ST+PST} model, and then allowing two overlapping circular components to emit at different temperatures, with the \texttt{ST+PDT} model.  
Only the most complex model, \texttt{PDT-U} allows for both hot spots to be described with such complexity. 
In this project, for the first time within the \NICER \ac{PPM} analyses, we consider also the emission from the remaining portion of the \ac{NS} surface, introducing the {\it elsewhere temperature} ($T_{\mathrm{else}}$), which can be added to all models. 

In Table \ref{tab:params}, we briefly list all the parameters used in this work to describe the \texttt{ST-U} ($+T_{\mathrm{else}}$) model. Within our naming convention, the \texttt{-U}, for {\it unshared}, signifies that the values of model parameters describing one hot spot are completely independent from the other \footnote{With the exception of the imposed condition that they do not overlap.}. Subscripts {\it p} and {\it s} are used to mark when a parameter refers to the primary or the secondary hot spot. 
The parameters in this table are in general common to all the models used in this paper; if two components are used to describe a hot spot, then the quantities below refer to either the only emitting spherical cap involved or - if both radiate - to the superseding one.  The only exception is the phase $\phi$, which refers to the masking component, if present. \\
\begin{deluxetable}{c|c}[b]
\label{tab:params}
\tablecaption{{Shared Model Parameters}}
\tablehead{\colhead{\bf{Symbol}} & \colhead{\bf{Meaning}}} 
\startdata
$M\,[\mathrm{M_\odot}]$& Mass of the \ac{MSP}\\
\hline
$R_{\mathrm{eq}}\,[\mathrm{km}]$ & Equatorial radius$^\dag$ of the \ac{MSP} \\
\hline
$D\,[\mathrm{kpc}]$ & Distance of the \ac{MSP}\\
\hline
$\cos(i)$ & cos of the inclination angle, the angle\\
& between the spin axis and line of sight\\
\hline
$N_{\mathrm{H}}\,[\mathrm{cm^{-2}}]$ & Hydrogen column density\\
\hline
$\log_{10}(T_p/\mathrm{K})$ & log of the primary hot spot temperature\\
\hline
$\log_{10}(T_s/\mathrm{K})$ & log of the secondary hot spot temperature\\
\hline
$\zeta_p \,[\mathrm{rad}]$ & Angular radius of the primary hot spot\\
\hline
$\zeta_s \,[\mathrm{rad}]$ & Angular radius of the secondary hot spot\\
\hline
$\theta_p \,[\mathrm{rad}]$ & Colatitude of the primary hot spot\\
\hline
$\theta_s \,[\mathrm{rad}]$ & Colatitude of the secondary hot spot\\
\hline
$\phi_p \,[\mathrm{cycles}]$ & Phase shift of the center of the primary \\
&hot spot compared to the reference\\
& phase set by the data\\
\hline
$\phi_s \,[\mathrm{cycles}]$ & Phase shift of the center of the secondary\\
&hot spot compared to the reference\\
& phase set by the data plus half a cycle\\
\hline
$\alpha$ & (\NICER, $\alpha_{\mathrm{XTI}}$, or \xmm, $\alpha_{\mathrm{MOS}}$) \\ 
& effective
 area \\
& energy-independent scaling factor\\
\hline
$\beta \,[\mathrm{kpc^{-2}}]$ & distance scaled effective area parameter $\alpha/D^2$\\
& (used as an alternative to $\alpha$ and $D$)\\
\hline
$\log_{10}(T_{\mathrm{else}}/\mathrm{K})$ & log of the temperature of the remaining\\
& surface of the \ac{MSP} \\
&(only adopted in few analyses) 
\enddata
\tablecomments{
Definition of the parameters that are common to all models adopted in this work. More detailed descriptions can be found in Section 2.3.3 of \citetalias{Vinciguerra2023a}. \\
$^\dag$ Hereafter also more simply referred to as radius.}
\end{deluxetable}
When a hot spot is modeled by two spherical caps, we need to define a few more parameters. We divide them into two cases: \texttt{PST} ({\it Protruding Single Temperature}) if only one component is emitting and \texttt{PDT} ({\it Protruding Double Temperature}) if both of them are. 
We describe all these parameters in Table \ref{tab:add_params}. A more detailed explanation of the adopted parametrization in models using {\texttt{PST}} and {\texttt{PDT}} hot spots can be found in Section 2.5.6 of \citetalias{Riley2019}.\\
\begin{deluxetable}{c|c|c|c}[b]
\tablecaption{{Additional Model Parameters for \texttt{PST} and \texttt{PDT} Hot Spots}}
\tablehead{\colhead{\bf{Symbol}} & \colhead{\bf{Meaning}} &\colhead{\texttt{PST}} & \colhead{\texttt{PDT}} } 
\startdata
$\zeta_{o/c}\,[\mathrm{rad}]$&Radius of the masking &\checkmark &\checkmark\\
&/ceding region& &\\
\hline
$\theta_{o/c}\,[\mathrm{rad}]$&Colatitude of the masking&\checkmark &\checkmark\\
&/ceding region&\\
\hline
$\chi\,[\mathrm{rad}]$& Azimuthal offset &\checkmark &\checkmark\\
& between the two caps$^*$ \\
\hline
$\log_{10} (T_{c}/1K)$& Temperature in log of & - &\checkmark\\
& the ceding region& &\\
\enddata
\tablecomments{
Definition of the parameters necessary to describe \texttt{PST} and \texttt{PDT} hot spots. 
A \texttt{PST} hot spot is characterised by an omitting and an emitting spherical cap, respectively labeled above with the subscript $_o$ and $_e$.
A \texttt{PDT} hot spot is characterised by a ceding and a superseding spherical cap, respectively labeled with the subscript $_c$ and $_s$. 
The parameters mentioned in the table with subscripts $_{o/c}$ hence refer to properties of the omitting component, for a \texttt{PST} hot spot, or to the ceding one, for a \texttt{PDT} hot spot. 
In presence of two hot spots, additional (initial) subscripts {\it p} or {\it s} will indicate whether the parameter refers to  the primary or the secondary hot spot. 
For \texttt{PST} parameters, a more detailed description can be found in Section 2.3.4 of \citetalias{Vinciguerra2023a}.\\
$^*$ $\chi/2\pi = \phi_{o/c} - \phi_{e/s}$, where $\phi$ marks phases and the subscripts clarify the corresponding component. 
}
\label{tab:add_params}
\end{deluxetable}
In this work, unlike in \citetalias{Riley2019} and \citetalias{Vinciguerra2023a}, we coherently incorporate \xmm data into   some of our analyses. 
\xmm is an imaging X-ray telescope. 
For this reason it performs better in identifying the photons that are generated by a particular source. 
Introducing this data set, with associated background, and trying to simultaneously fit both \NICER and \xmm data, 
allows us to better evaluate which fraction of the registered events (for both data sets) actually come from \joo and which are due to external contributions (see Section 4.2 of \citealt{Riley2021} for more details). 
We describe below how we model uncertainties in the instrument responses of both \NICER and \xmm.\\

{\bf Modeling of the instrument response}: 
    When we analyze only the \NICER data set B19v006, we use the same parametrization described in Section 2.2 of \citetalias{Vinciguerra2023a} and summarised below. 
    To model uncertainties in the instrument response, we adopt a variable $\beta\,[\mathrm{kpc^{-2}}] = \alpha_{\mathrm{XTI}} D^{-2}$, where $D$ is the pulsar distance in $\mathrm{kpc}$ and $\alpha_{\mathrm{XTI}}$ the energy independent factor that scales the reference response matrix, and hence the total effective area, of the \NICER X-ray Timing Instrument (XTI). 
    $\beta$ represents the only combination of $\alpha_{\mathrm{XTI}}$ and $D$ to which our analysis is sensitive. 
    When both \xmm and \NICER data sets are used, we instead sample from $\alpha_{\mathrm{XTI}}$, $\alpha_{\mathrm{MOS}}$ and $D$. 
    As described in Table \ref{tab:params}, 
    $\alpha_{\mathrm{MOS}}$ represents 
    the energy-independent scaling factor applied to the reference instrument responses 
    of \xmm two cameras: MOS1 and MOS2 (i.e. we assume 
    $\alpha_{\mathrm{MOS}}=\alpha_{\mathrm{MOS1}}= \alpha_{\mathrm{MOS2}}$, which may not actually be the case). 
    We make this parametrization choice to describe the correlation between the \NICER and \xmm instrument responses as explained in see Section 2.4 of \citet[][]{Riley2021}. As in \citet{Salmi2022}, in this work we adopt the {\it compressed effective area prior} \citep[see Section 4.2 of ][]{Riley2021}. 
    
    In this paper we chose to apply pretty conservative priors for the $\alpha$ parameters. These are comparable to what was used in  \citetalias{Riley2019} and more constraining compared to what was adopted for the headline results of \citet{Riley2021}, but still broad enough to allow follow-up tests, adopting importance sampling, with more restricted priors, which could reflect more accurately the current understanding of \NICER and \xmm calibration uncertainties. 
    
\subsubsection{Priors and Settings}
\label{subsubsec:ParamPrior}
For this paper we use the same priors and settings introduced in Section 2.2 of \citetalias{Vinciguerra2023a}. 
These include a flat prior in the joint mass and radius parameter space, where hard boundaries are applied: the mass is constrained to be in the range $\in (1.0,3.0)\, \mathrm{M_\odot}$; while the radius must be less then 16\,km. Similarly to \citetalias{Riley2019,Vinciguerra2023a}, \citet{Salmi2022,Salmi2023} we also limit the compactness and the surface gravity. 
Differently from \citetalias{Riley2019}, we adopt the instrumental PI channels in the range $[30,300)$ and we consider isotropic priors (flat in cosine) for the inclination angle $i$ and the colatitudes of the hot spots' centers $\theta_p$ and $\theta_s$.  \\
Priors describing the additional emitting component in the {\texttt{ST+PDT}} and {\texttt{PDT-U}} models are constrained by the overlapping condition, which imposes overlaps between the superseding and the ceding spherical caps. These priors are therefore coupled to the value of the parameters describing shape and location of the super\-ceding component. 
These dependencies are similarly implemented as for the {\texttt{ST+PST}} model, starting from the angular radius of the ceding region $\zeta_c$ (see Section 2.5.6 of \citetalias{Riley2019} for more details). 
When used, the temperature of the additional elsewhere component $\log_{10}\left(T_{\mathrm{else}}/\mathrm{K}\right)$ is sampled uniformly between the bounds $(5.0, 6.5)$. \\
In this work, as well as in \citetalias{Vinciguerra2023a}, we use both {\it high-resolution} (HR) and {\it low-resolution} (LR) \XPSI settings (the specific parameters, that define such settings, and their values are presented in Section 2.3.1 of \citetalias{Vinciguerra2023a}). 
The latter allows us to run complex models, whilst saving significant computational resources.  According to the results presented in \citetalias{Vinciguerra2023a}, no significant changes are expected in the results adopting the low-resolution settings. Below we test whether this hypothesis holds also in the case of the real \NICER data. 

\subsubsection{MultiNest}
\label{subsec:multinest}
\XPSI needs to be coupled with a sampling program. In this work we use PyMultiNest \citep{Buchner2014}, a library that allows us to interface easily with \MultiNest. \MultiNest is a Bayesian inference technique and a program that targets the estimation of the evidence. 
Doing so, it explores the parameter space, and so allows for parameter estimation. 
More details are given in Section 2.4 of \citetalias{Vinciguerra2023a}; in  Table \ref{tab:MultiNestParams} we summarize the most relevant settings for the analyses performed in this work. 
Our default starting \MultiNest settings are: sampling efficiency 0.3; evidence tolerance 0.1; live points 1000 and multi-mode/mode-separation off (SE 0.3, ET 0.1, LP 1000, MM off). 
The sampling efficiency initially set is then modified within \XPSI to account for the effective unit hyper-cube volume of the prior space considered in the analysis. 

\begin{deluxetable*}{c|c|c|c}
\tablecaption{{Most Relevant Parameters Describing \MultiNest Settings}}
\tablehead{\colhead{\bf{Symbol}} & \colhead{\bf{Meaning}} &\colhead{\bf{Range or Scale}}&\colhead{\bf{Our default}}} 
\startdata
\hline
SE & Sampling Efficiency & Typically $\in (0.1-1)$ & 0.3\\
\hline
ET & Evidence Tolerance & Typically $\le0.5$ & 0.1\\
\hline 
LP & Live Points & order hundreds to thousands & 1000\\
\hline
MM & Multi-mode/mode-separation & on/off & off\\
\enddata
\tablecomments{
The term `typically' is used to denote values that we have encountered in the literature.  By `Our default' we mean the default values used in our analysis unless otherwise specified.  The \MultiNest advised values are 0.5 for evidence tolerance and 0.3 or 0.8 for sampling efficiency, respectively if the main goal is evidence calculation or parameter estimation \citep{MultiNest_2008,BuchnerGitHubMN}. 
No formal number of live points have been suggested, likely because an adequate number would depend on the problem at hand. 
Given the same settings, adopting the mode-separation modality slightly worsens the precision of the final results, as live points no longer move as efficiently. 
Lower numbers for sampling efficiency and evidence tolerance, and higher numbers of live points, imply higher accuracy in the calculation; however they also significantly increase the computational resources required. 
}
\label{tab:MultiNestParams}
\end{deluxetable*}

\subsection{Test Cases}
\label{subsec:runs}
The main goal of this paper is to establish a baseline for the analysis of the upcoming new \joo data sets and their interpretation. It is supported by the simulations carried out in \citetalias[][]{Vinciguerra2023a}. 
To achieve this goal we first set up a few exploratory runs, aiming to test the robustness of the solutions found by \citetalias{Riley2019}. 
We first try to reproduce those results, with the new set-up described in this Section. Then we investigate the 
effect of background constraints, by including the \xmm data set in our inference runs.  We look at the possible presence of multi-modal structures in the posterior surface, as well as conceivable shortcomings of the analysis (in light of what was found in \citetalias{Vinciguerra2023a}). 

Given the scope of this paper, we decided to focus most of our tests on the two simplest models \texttt{ST-U} and \texttt{ST+PST}. The latter was identified as the preferred model in \citetalias{Riley2019} (and associated with the headline mass-radius result); the former was found to be the simplest model that could represent the \joo\ \NICER data without showing clear structures in the residuals. With these two models we test robustness of the obtained inferred parameter values and their dependencies on random sampling processes and analysis settings. 
We also begin to explore the parameter space of the more and most complex two hot spot models: \texttt{ST+PDT} and \texttt{PDT-U}. 
 Since \citetalias{Vinciguerra2023a} 
uncovered the presence of prominent multi-modal structures in the posterior surface, for \NICER-only analyses, we here perform one inference adopting the mode-separation variant with $10^4$ live points for each model; these are our (\NICER-only) reference runs. 
 These settings are proven (see Section \ref{subsubsec:STU_ImpactSettings}) to generate stable posteriors when the \texttt{ST-U} model is used; however this is not necessarily the case for the more complex models, for which a more detailed testing would have been computationally too expensive.
For all four models, we also perform a preliminary, production run including \xmm data in the analysis. 

We summarize all of the runs carried out, and their settings, in Table \ref{tab:runs}.

\begin{table*}[t]
\begin{tabular}{l|l|l|l|l|l|l|l}
{\bf Model} & {\bf X-PSI settings} &{\bf SE} & {\bf ET} & {\bf LP} & {\bf MM} & {\bf \xmm} & {\bf N}\\
\hline
\hline
\texttt{ST-U} & HR & 0.3 & 0.1 & $10^3$ & off & no & 3\\
& HR & 0.3 & 0.1 & $3\times 10^3$ & off & no & 1\\
& HR & 0.3 & 0.1 & $6\times 10^3$ & off & no & 1\\
& HR & 0.3 & 0.1 & $10^4$ & off & no & 1\\
& HR & 0.1 & 0.1 & $10^3$ & off & no & 1\\
& HR & 0.8 & 0.1 & $10^3$ & off & no & 1\\
& HR & 0.3 & 0.001 & $10^3$ & off & no & 3\\
&{\bf HR }&{\bf 0.3} &{\bf 0.1 }& $\mathbf{10^4}$& {\bf on }&{\bf no }&{\bf 1}\\
& HR & 0.3 & 0.1& ${10^4}$ & on & yes & 1\\
\texttt{ST-U}+$T_{\mathrm{else}}$ & HR & 0.3 & 0.1 & $10^3$ & off & no & 1\\
\hline 
\texttt{ST+PST} & HR & 0.3$^*$ & 0.1 & $10^3$ & off & no & 3 \\
& LR & 0.3 & 0.1 & $10^3$ & off & no & 2 \\
& HR & 0.3 & 0.1 & $10^3$ & on & no & 1 \\
$^{**}$& LR & 0.3 & 0.1 & $10^3$ & off & no & 1 \\
$^{**}$&{\bf LR }&{\bf 0.3 }&{\bf 0.1 }& $\mathbf{10^4}$ &{\bf on} &{\bf no }&{\bf 1 }\\
$^{**}$& LR & 0.8 & 0.1 & ${10^3}$ & off & yes & 1 \\
\texttt{ST+PST}+$T_{\mathrm{else}}^{**}$& LR & 0.3 & 0.1 & $10^3$ & off & no & 1\\
\hline 
\texttt{ST+PDT} & LR & 0.8 & 0.1 & $10^3$ & off & no & 1\\
& LR & 0.8 & 0.1 & $10^3$ & on & no & 1\\
& {\bf LR }&{\bf  0.8 }&{\bf  0.1 }& $\mathbf{10^4}$ &{\bf  on} &{\bf  no} &{\bf  1}\\
& LR & 0.8 & 0.1 & ${10^3}$ & off & yes & 1\\
\hline 
\texttt{PDT-U} & LR & 0.8 & 0.1 & $10^3$ & off & no & 1\\
 &{\bf  LR }&{\bf  0.8 }&{\bf  0.1 }& $\mathbf{10^4}$ &{\bf  on }&{\bf  no }&{\bf  1}\\
& LR & 0.8 & 0.1& ${10^3}$ & off & yes & 1\\
\end{tabular}

\caption{
Collection of inference runs performed for this work and the corresponding properties and settings. 
The first column represents the name of the hot spot model adopted for the run (see Section \ref{subsubsec:HSmodels} and Section 2.5 of \citetalias{Riley2019}). When the conventional model name is followed by $T_{\mathrm{else}}$, the run also includes the modeling of the temperature of the rest of the \ac{MSP}'s surface, exterior to the two hot spots. The second column describes whether the run was performed with high (HR) or low (LR) resolution for some of the main \XPSI parameters. The four columns that follow represent the \MultiNest settings as described in Table \ref{tab:MultiNestParams}. The seventh column is used to flag inference runs where \NICER and \xmm data have been fitted at the same time; and the last column shows how many identical repetitions of the same run were performed to test the impact of random sampling processes. Reference runs for this study are highlighted in bold.\\
$^*$: Two of these three runs were resumed with sampling efficiency 0.8, as was done in \citetalias{Riley2019};\\
$^{**}$: These runs were performed adopting the updated CoH prior.\\
}
\label{tab:runs}
\end{table*}

\section{Data Sets}
\label{sec:dataset}
\subsection{NICER B19v006 Data Set}

For this work, we use the same dataset as in the initial \NICER analyses of \joo \citep{Miller2019,Riley2019}, with data processing as described in \citet{Bogdanov19a} (B19).
The dataset contains 1.936 Ms of exposure time collected over the period 2017 July 24 to 2018 December 9. However we have recalibrated the gain using the \texttt{nicerpi} tool, updating the calibration to use the gain calibration file \texttt{20170601v006}. The response matrices matched to this gain solution are in \NICER CALDB file \texttt{xti20200722}. 
{Compared to the inferences reported in \citetalias{Riley2019} (which also included energy channels $25-30$), analyses of this recalibrated data set need to be limited to energy channels $\ge 30$, as the procedure adopted to create the updated B19v006 data set from the original B19 one allows for 
possible calibration errors at these excluded lowest channels. 
In Section \ref{subsec:STU_comparisonR19} we demonstrate that the effects of removing these channels are minor. This is in agreement with the analysis of \citet{Miller2019}, which considered only channels $\ge 40$ and found that this choice did not affect their results. }

\subsection{XMM Data Set}
\label{subsec:dataset_XMM}
The \xmm data of \joo used in this analysis (first presented in \citealt{BogdanovGrindlay2009}) are from two archival observations obtained on 2001 June 19 (ObsID 0112320101) and 2007 December 12 (ObsID 0502290101). We only consider the EPIC MOS1 and MOS2 `Full Frame' mode imaging observations; although they do not possess adequate sampling time to resolve pulsations from \joo they provide reliable low background phase-averaged source spectra.  The EPIC pn data acquired in `Timing' mode, which provides imaging only in one direction of the detector, is not used due to the considerably larger uncertainties in the calibration of the instrument in this observing mode. The data reduction and extraction of the EPIC MOS data were carried out using the Science Analysis Software (SAS\footnote{The \xmm SAS is developed and maintained by the Science Operations Centre at the European Space Astronomy Centre and the Survey Science Centre at the University of Leicester.}) version \texttt{xmmsas\_20211130\_0941}. The event data were screened for instances of high background count rates and the recommended PATTERN ($\le$12 for MOS1/2) and FLAG ($0$) filters were applied. This resulted in $121.2$\,ks and $120.9$\,ks of total effective exposure for MOS1 and MOS2, respectively, from combining both observations.

\section{Results}
\label{sec:results}
In the following, we present the results of the inference runs described in Section \ref{subsec:runs} and summarized in Table \ref{tab:runs}. 
We consider each hot spot model presented in Section~\ref{subsubsec:HSmodels}, going from the simplest to the most complex. 
For the simplest, {\texttt{ST-U}} and {\texttt{ST+PST}}, we start with parameter estimation analyses that allow an easier comparison with the findings of \citetalias{Riley2019} and explore our sensitivity to settings and random sampling processes, as in \citetalias{Vinciguerra2023a}. 
The main data and post-processing routines necessary to reproduce our main results are available at \citet[][the Zenodo link will be made public, and the files available, once the publication is accepted]{J0030zenodoBRAVO}.

As estimating mass and radius of \acp{MSP} is the main science goal of \NICER, in this section we focus in particular on the inferred posterior distributions of these two parameters. We also show posteriors for compactness, since our analysis is expected to be most sensitive to this particular combination of mass and radius\footnote{When computing compactness as defined in Section \ref{sec:method} we use the equatorial radius $R_{\mathrm{eq}}$.}. 
The corner plots reported in most of the figures that follow show the 1D and 2D posterior distributions of these three parameters. 
As in \citetalias{Vinciguerra2023a}, \citetalias{Riley2019}, \citet{Riley2021, Salmi2022}, the colored band present in the 1D posterior plots shows the 
area enclosed within the $\sim16\%$ and $\sim 84\%$ quantiles of the 1D marginalized posterior distributions (medians and corresponding lower and upper limits are written on top of each 1D posterior); while the contours in the 2D marginalized distributions represent the $\sim 68.3\%, 95.5\%$ and $99.7\%$ credible regions.
As explained in Section 4 of \citetalias{Vinciguerra2023a}, the \XPSI post-processing tools adopt the GetDist\footnote{\url{https://getdist.readthedocs.io}} KDE to smooth the distribution of the posterior samples. 
In 2D posterior plots, this may introduce artificial gradients in the density, 
when sharp cut offs (more complex than a threshold on a single parameter) are present in the prior. 
This is for example the case for the 2D posterior density plot of radius and compactness (see also footnote 12 of \citetalias{Vinciguerra2023a}). 

 In this section, as well as in the discussion (Section \ref{sec:discussion}), for simplicity in guiding our thoughts, we sometimes refer to a single sample, the maximum likelihood of the particular run or mode that we are considering. 
We use it as a reference point to describe some of the properties of such portion of the posterior. However, generalizations need to be carefully evaluated since a single sample cannot fully represent the whole range of possibilities spanned by a posterior mode\footnote{In fact, as in \citetalias{Vinciguerra2023a}, in one instance we use the maximum posterior sample to highlight spread in the parameter space corresponding to a single posterior mode.}. 

\subsection{\texttt{ST-U}}
\subsubsection{Reproducing R19 Results}
\label{subsubsec:STU-R19}
\citetalias{Riley2019} identified the \texttt{ST-U} as the simplest model that could represent the original \NICER data set of \joo analyzed in that paper. 
The derived 68\% credible interval for mass and radius, assuming \texttt{ST-U}, were respectively $1.09^{+0.11}_{-0.07}\,\mathrm{M_\odot}$ 
and $10.44^{+1.10}_{-0.86}$\,~km.
For that inference, the \MultiNest settings adopted to analyse \joo were: 
 SE 0.3, ET 0.001, LP 1000, MM off (hereafter we refer to these as the \citetalias{Riley2019} \MultiNest settings for \texttt{ST-U})\footnote{One of the three runs was done with MM set to be on and ET 0.1. However, since the results obtained were very similar to the other two runs, we will hereafter neglect it.}, with the posteriors of the 3 tested runs, overlapping (see e.g. the posteriors of run 1 in the left panel of Figure \ref{fig:STU_res}). 
Within this new framework (which, compared to the analyses of \citetalias{Riley2019}, incorporates an updated \NICER data set -adapted to the most recent \NICER instrument response- as well as upgraded software), 
we therefore perform three identical and independent analyses using the same \MultiNest settings. The obtained posterior distributions for mass, radius and compactness are reported in the left corner plot of Figure \ref{fig:STU_res}. 
The variability of the results inferred from these three runs suggests that, in contrast to what was reported in \citetalias{Riley2019}, in our current analysis framework (including all the changes previously described) these settings 
are insufficient to exhaustively explore the parameter space. For this reason, in the same plot, we also report posterior distributions for the SE 0.3, ET 0.1, LP $10^4$, MM on, HR run, which we consider effectively representative of the \texttt{ST-U} results in this new framework (see Figure \ref{fig:STU_settings}). Although we notice a slight increase in the inferred median value of mass and radius as well as in the widths of their posteriors, they are in good agreement with the findings of \citetalias{Riley2019}. 
If instead of focusing on inference runs with $10^4$ live points, we focus on the results 
derived here with \MultiNest settings similar to those adopted in \citetalias{Riley2019}, the newly inferred radius and mass posterior distributions are narrower and display considerably larger medians (differences in radius can be $\gtrsim 1$\, km). 
We consider the LP=$10^4$ run to be more robust since it explores the parameter space better and is more stable (see indeed the overlap between posteriors with $\mathrm{LP} \ge 6\times 10^3$ reported in Figure \ref{fig:STU_settings} and discussed in Section \ref{subsubsec:STU_ImpactSettings}). 

The right corner plot of Figure \ref{fig:STU_res} displays the posterior distributions obtained with our default \MultiNest settings (SE 0.3, ET 0.1, LP $10^3$, MM off). Not surprisingly, also in this case, where the accuracy requirement over the evidence estimate is lower, the results exhibit some variability.  
In addition, in the same plot,  we report the findings obtained adopting the \texttt{ST-U}$+T_{\mathrm{else}}$, which includes the possibility that the whole surface of the \ac{NS} could emit in the \NICER sensitivity band. We find that adding this further element in the model produces results consistent with the other inference runs performed with the same settings.

{
    \begin{figure*}[t!]
    \centering
    \includegraphics[
    width=18cm]{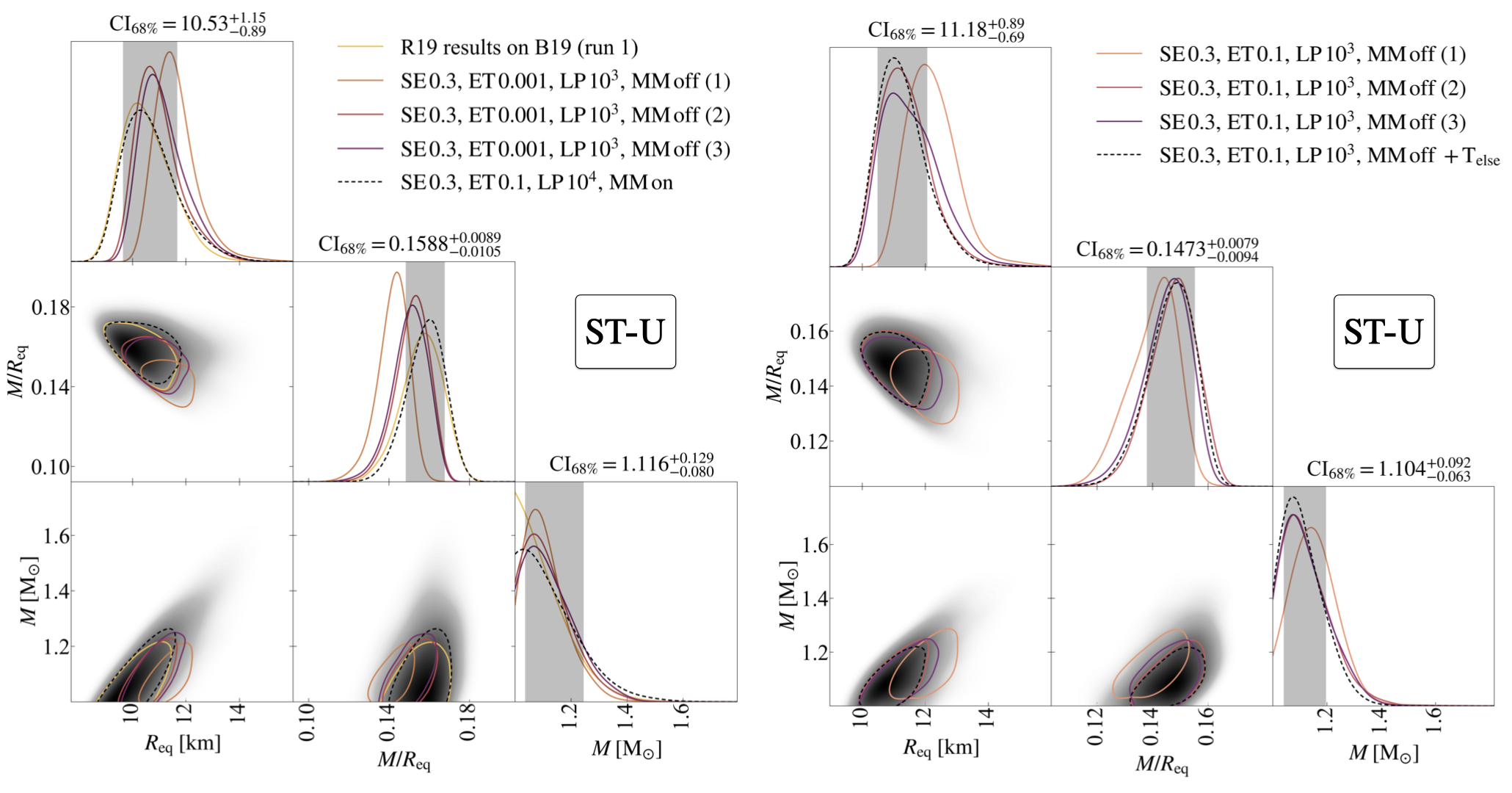}
    \caption{\small{ 
    \texttt{ST-U} posterior distributions (smoothed by GetDist KDEs) from 9 inference runs, for radius, compactness and mass. 
    The corner plot on the left shows results for three inference runs with \MultiNest settings identical to those adopted by \citetalias{Riley2019}, but obtained in the new \XPSI and data framework described in Sections \ref{sec:method}, \ref{sec:dataset} (differences listed in Section \ref{subsec:STU_comparisonR19}). For all these runs we adopted the high-resolution \XPSI settings. For comparison, in solid yellow lines we also show the distributions found for run 1, \texttt{ST-U} model, in \citetalias{Riley2019}.  
    In dashed black we present posterior distributions for the SE 0.3, ET 0.1, LP $10^4$, MM on run, which we consider to be the reference runs for \texttt{ST-U} model in this work.
    The corner plot on the right shows the results for three inference runs with our default \MultiNest settings. 
    We also report posterior distributions for the run allowing the whole surface of the \ac{MSP} to emit (labeled as \texttt{ST-U}$+T_{\mathrm{else}}$). 
    Each of the two corner plots shows new posteriors from four inference runs which use different \MultiNest settings, as reported in the legend (for definitions see Section \ref{subsec:multinest}). 
     Curves in the representation of the 2-dimensional posteriors trace the 68\% credible area.
    On top of the 1D posteriors, we report the 68\% credible intervals  (representing the area within the 16\% and 84\% quantiles in the 1D marginalized posterior) starting from the median of the distributions. 
    These values, as well as the colored areas, refer to the two dashed black lines: the run enabling the mode-separation modality, in the left corner plot, and the run adopting the \texttt{ST-U}$+T_{\mathrm{else}}$ model, in the right corner plot. 
    } 
    }
    \label{fig:STU_res}
    \end{figure*}
}
\subsubsection{Impact of Analysis Settings}
\label{subsubsec:STU_ImpactSettings}
The findings of \citetalias{Vinciguerra2023a} highlight the importance of \MultiNest settings. For \texttt{ST-U} (computationally the cheapest of our models) we therefore tested the impact of adopting different values of sampling efficiency or evidence tolerance, starting from our default settings (based on the settings adopted for models more complex than \texttt{ST-U} in \citetalias{Riley2019}).
The inferred posterior distributions are reported in the left corner plot of Figure \ref{fig:STU_settings}. 
We report: in pink, the results for all the three runs with our default settings, and in brown those for the three inference runs, performed with ET 0.001, shown individually in the right and left plot respectively of Figure \ref{fig:STU_res}. 

On the right corner plot, we demonstrate the importance of setting an adequate number of live points to exhaustively cover the model parameter space and derive a stable solution. 
Again, in pink, we represent parameter estimates for three inference runs adopting our default \MultiNest settings. We then show posterior distributions obtained respectively with runs adopting 3k, 6k, and 10k live points, demonstrating the gradual convergence to slightly lower values of radii and larger uncertainties. 
These results demonstrate that in this setup (and having fixed the sampling efficiency to 0.3 and the evidence tolerance to 0.1) we need a minimum number of live points somewhere in the range $(3-6)\times 10^3$
to guarantee an adequate exploration of the model parameter space. 

In both the corner plots, as a reference, we also plot the posterior distributions obtained by the inference run SE 0.3, ET 0.1, LP $10^4$, MM on, HR. 
Adopting the latter as a reference, caveat the low number statistics, these results hint at the following conclusions:
\begin{itemize}
    \item In this new framework, $(3\times)\, 10^3$ live points are not enough to exhaustively explore the model parameter space; 
    \item About 1 in every 3 runs seems to give a significantly different median value for the radius posterior, both with our default \MultiNest settings and with the variation on the evidence tolerance (ET 0.001); 
    \item Compared to when we adopt ET 0.001, our default \MultiNest settings (with ET 0.1) seem to lead to a greater variation in the radius and mass medians (but smaller in the compactness); these medians are also in general further away from the values identified with the reference SE 0.3, ET 0.1, LP $10^4$, MM on, HR inference run; 
    \item However, our default settings seem to also recover wider posteriors compared to their ET 0.001 counterparts; wider posteriors, again compared to the ET 0.001 runs, are also recovered 
    with the \texttt{ST-U} reference run; 
    \item Changing the value of sampling efficiency does not seem to significantly affect the median value of the posteriors; 
    however, when lowering it to 0.1, the widths of the posterior distributions increase, approaching the values obtained if one adopts a higher number of live points;
    \item Given the appearance of a bias towards higher radius values when $10^3$ live points are used, multiple runs with this setting seem unlikely to deliver the same result as one would obtain from a single run with a higher number of live points. 
\end{itemize}

{
    \begin{figure*}[t!]
    \centering
    \includegraphics[
    width=18cm]{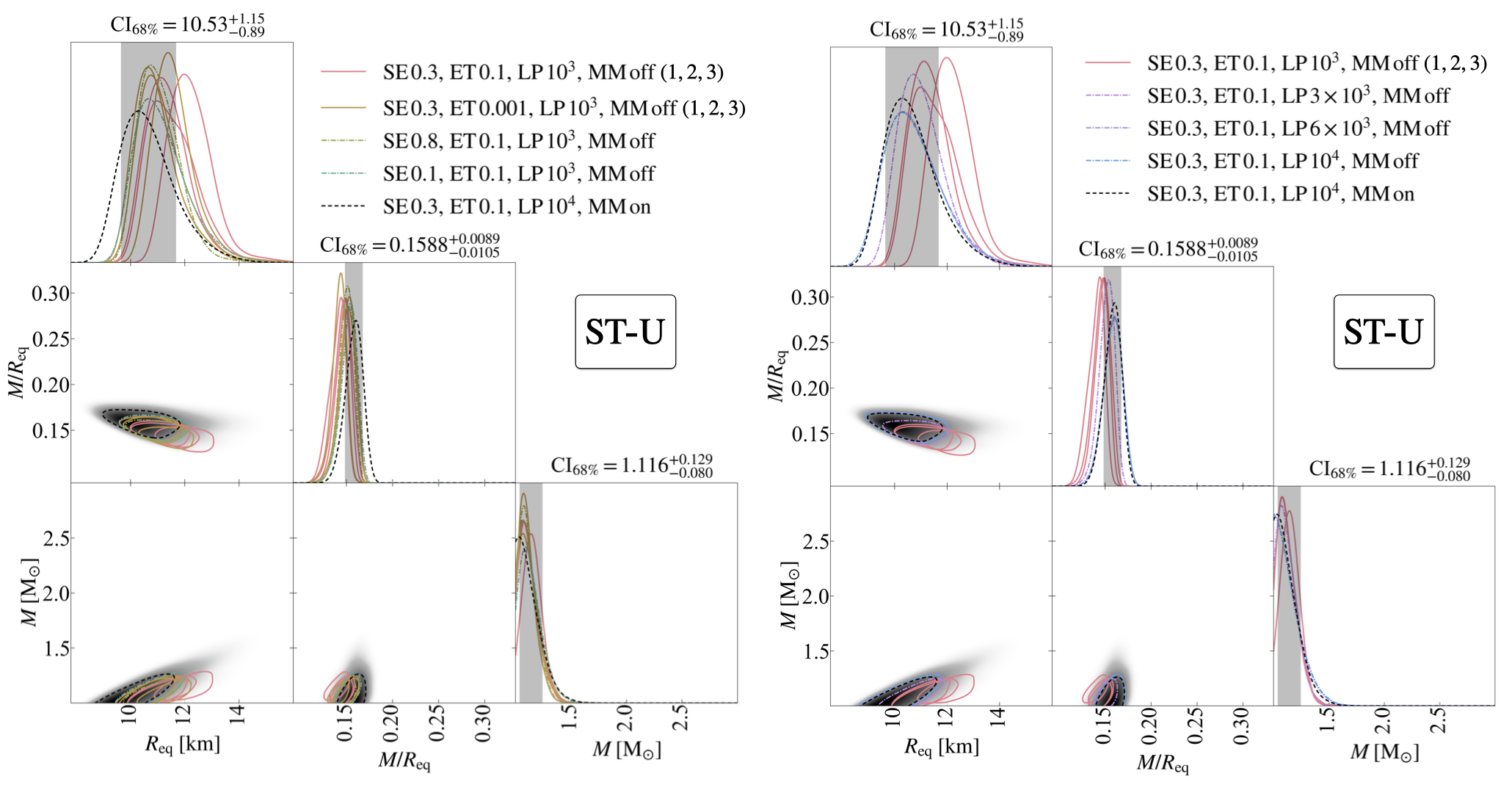}
    \caption{\small{\texttt{ST-U} posterior distributions (smoothed by GetDist KDEs) from 16 runs, for radius, compactness, and mass. This figure highlights the effects of adopting different \MultiNest settings (for all these runs we adopted the high-resolution \XPSI settings). On the left plot we show results for 9 \texttt{ST-U} inference runs: they illustrate, for the specific problem at hand, what is the effect of changing evidence tolerance or sampling efficiency, compared to our default settings. 
    The corner plot on the right instead shows the effect of increasing the number of live points.  
    As a reference, in both corner plots, we also report with dashed black lines the posterior distributions obtained with the SE 0.3, ET 0.1, LP $10^4$, MM on run; numbers on top of the 1D posteriors, as well as shaded areas, refer to this inference run. 
    See caption of Figure \ref{fig:STU_res} for further details. 
    } 
    }
    \label{fig:STU_settings}
    \end{figure*}
}
\subsubsection{Model Exploration}
The different mass and radius values obtained for the sequence of nested surface pattern models analyzed in \citetalias{Riley2019} already suggested the presence of multiple modes in the likelihood surface. 
In this work, we highlight and expand on these findings. 
We focus here on the posterior distributions obtained with the  SE 0.3, ET 0.1, LP $10^4$, MM on, HR \texttt{ST-U} inference run. 
Here we find three modes in the posterior, with considerably different inclinations and hot spot locations (configuration corresponding to the main one reported in panel A of Figure \ref{fig:config}).  
Interestingly, they show the same hot spot configuration modes as found \citetalias{Vinciguerra2023a} when analysing with the \texttt{ST-U} model a synthetic data set produced with the \texttt{ST+PST} model (see Figure 7 of \citetalias{Vinciguerra2023a} for a visual representation). 
As in that case, the two secondary solutions have similar likelihood and local evidence values, but they perform considerably worse than the main identified mode (the maximum likelihood geometry, for the equivalent run but with MM off, is reported in panel A of Figure \ref{fig:config}). 
They also differ considerably from it in inferred mass and radius 
(means and standard deviations associated with them are reported in Table \ref{tab:STU_MM}). 

\begin{table*}[tbp]
\hspace*{-1cm}
\begin{tabular}{l|l|l|l|l|l}
 &  \multicolumn{3}{c|}{{\bf NICER}} & \multicolumn{2}{c}{{\bf NICER \& XMM}} \\

 &  {\bf Mode 1} &  {\bf Mode 2} &  {\bf Mode 3} &  {\bf Mode 1} &  {\bf Mode 2} \\
\hline
${\mathbf{R_{\mathrm{eq}}}}$ [km] & $10.7 \pm 1.0$ & $15.5 \pm 0.4$ & $15.5 \pm 0.4$& $14.9 \pm 1.0$ & $11.4 \pm 1.1$  \\
\hline
{\bf M} $\mathrm{[M_\odot]}$ &  $1.1\pm 0.1$ &$1.6\pm 0.1$ & $1.6\pm 0.1$& $1.9\pm 0.2$ & $1.4\pm 0.2$\\
\hline 
$\boldsymbol{{\mathbf{\mathrm{max}}}\left(\log\left(\mathcal{L}\right)\right)}$ & -35735 & -35758 & -35757 & -42661 & -42666\\
\hline 
$\boldsymbol{\log\left(\mathcal{E}\right)}$ & -35788 & -35810 & -35808 & -42714 & -42718\\
\hline 
{\bf Configuration} & panel A, Fig. \ref{fig:config} & middle panel & rightmost panel & panel C, Fig. \ref{fig:config}& Panel D, Fig. \ref{fig:config} \\
& & of Fig. 7, \citetalias{Vinciguerra2023a}& of Fig. 7, \citetalias{Vinciguerra2023a} & &\\
\end{tabular}
\caption{
Table describing the main properties of the three modes identified adopting the \texttt{ST-U} model. 
The first two rows show means and standard deviations of mass $M$ and equatorial radius $R_{\mathrm{eq}}$ posterior distributions; 
the last two the corresponding maximum log-likelihood and local log-evidence values. 
}
\label{tab:STU_MM}
\end{table*}
From the right corner plot of Figure \ref{fig:STU_settings}, we can compare the mass, radius, and compactness posterior distributions of this inference run to those obtained adopting the same \MultiNest settings but disabling the mode-separation. 
When we enable the mode-separation the posteriors are slightly narrower but otherwise they are very similar to each other: the 68\% credible intervals for the \joo radius, with and without the activation of the mode-separation, are respectively 
$10.53^{+1.15}_{-0.89}$\,km and $10.60^{+1.28}_{-0.96}$\,km; while for the mass they are $1.12^{+0.13}_{-0.08}\, \mathrm{M_\odot}$ and $1.12^{+0.15}_{-0.09}\, \mathrm{M_\odot}$. 
While the latter lets the live points freely and more optimally explore the parameter space, the former, our reference run for \texttt{ST-U} \NICER-only analyses, allows us to identify modes in the inferred posterior, even when not visible by eye. 
\subsubsection{Joint Analysis of NICER and XMM-Newton Data}
\label{subsubsec:STU_NxXMM}
In the left corner plot of Figure \ref{fig:STU_STPST_NxX}, we report the main findings from the joint \NICER and \xmm inference analysis. 
They refer to the SE 0.3, ET 0.1, LP $10^4$, MM on, HR run,  performed with the setup described in Section \ref{sec:method}. 
The configuration associated with the maximum likelihood sample from its posterior distribution is shown in panel C of Figure \ref{fig:config}. 

The 1D posteriors  for mass, radius, and compactness all move to larger values. 
In particular, the distribution clearly reaches the radius upper boundary of 16\,km. 
Because our analysis is particularly sensitive to the compactness, this radius upper limit is expected to  indirectly constrain the mass posterior as well. 
The plot suggests that the peak of the radius posterior would correspond to even larger radii than allowed by our (physically motivated) prior. 
There is, however, an elongated tail of the posterior involving considerably lower radii. 
The posteriors belonging to this tail are also characterised by considerably lower masses (see the 2D posterior distribution of mass and radius), in closer agreement with what was found as the main mode when only the \NICER data was analyzed. 
This tail in the radius posterior of the joint \NICER and \xmm inference corresponds to the secondary mode found by the sampler, whose configuration is reported in panel D of Figure \ref{fig:config}. The average and standard deviation of this secondary peak in the posterior, as well as of the main peak, are reported in the last two columns of Table \ref{tab:STU_MM}. 
Both modes show a compactness posterior distribution which peaks at considerably higher values compared to the results obtained when only \NICER data are analyzed. 
Including the \xmm data in the inference process limits the contribution of the background (see Figure \ref{fig:BKG}), compared to what is inferred from \NICER data alone. 
To offset the reduction in the unpulsed background, the inferred compactness increases, creating a larger unpulsed signal arising from the star
(although with the opposite effect, the same logic was applied to explain the findings from joint \NICER and \xmm analyses also in \citealt{Riley2021,Salmi2022} for \joh). 
The same trend emerges for both \texttt{ST-U} and \texttt{ST+PST} models. 

The higher radius and compactness values, yield lower angular radii describing the two hot spot sizes and allow the hot spot at higher colatitude (closest to the pole) to move slightly closer to the equator. 
The colatitudes of the hot spots are however not tightly constrained and this leads to a visible bi-modal structure in the posterior of the parameters describing the hot spot properties (see Figure \ref{fig:params}). 
This is due to the ambiguity of the primary and secondary labels associated with the hot spots 
\footnote{For models which adopt the same complexity to describe both hot spots, we label as primary the hot spot with lower colatitude. This implies that ambiguity on the hot spot associated with a specific label arises if there is a significant posterior mass (note that this is not the same as the posterior of the pulsar mass parameter, or the mass posterior). for similar colatitude values for both the hot spots (this chance increases with broader posteriors).}.
Since the \ac{MSP} radius inferred through the secondary mode peaks at values only marginally larger than the one found with \NICER-only data, the hot spots' angular radii still decrease, but not as much as for the main posterior mode (see Figure  \ref{fig:params}). 
The configurations arising from the main mode of  the joint \NICER and \xmm inference show temperatures and inclination that are very similar to those found in the main mode when analysing \NICER-only data. 
This is not the case for the secondary mode, which points to a more edge on configuration, with inclination compatible with zero. 
In this new setting, hot spots are located close to the equator and characterised by significantly lower temperatures. 
{
    \begin{figure*}[t!]
    \centering
    \includegraphics[
    width=18cm]{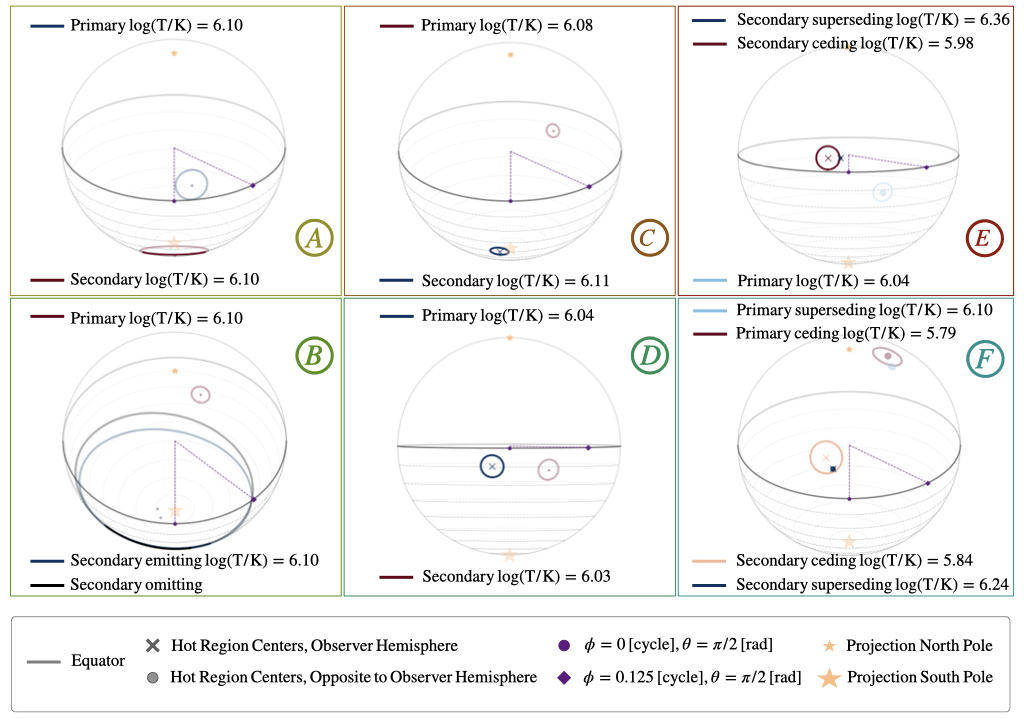}
    \caption{\small{Representation of the main geometrical modes found with the analyses presented in this study. 
    The surface patterns are shown from the observer point of view (i.e. according to the inferred inclination of the specific sample considered), at phase zero of the data. 
    Hot spots on the same (opposite) hemisphere compared to the observer are plotted with full (dimmed) colors and and a cross (circular) marker at the centre. 
    The color of the hot spot components changes from blue to red from the highest to the lowest temperature of the considered surface pattern (with higher precision compared to the temperature reported in the legend). Black is used for masking components. 
    In all cases, we show, as a reference point, the hot spot geometry associated with the maximum likelihood sample of that specific mode within the considered inference run.  Although this single sample cannot capture the whole possible variations present in a mode, we hope in this way to provide a simplified but more concrete indication of the configurations belonging to the mode.
    Panels A and B refer to the configurations associated with the main mode of \texttt{ST-U} (SE 0.3, ET 0.1, LP $10^4$, MM off, HR) and \texttt{ST+PST} (SE 0.3, ER 0.1, LP $10^4$, MM on, LR) models, \NICER-only analyses. Panels C, D, E and F refer to the configurations found in joint \NICER and \xmm inferences. 
    Panels C and D are representative of the main (C) and secondary (D)  modes derived adopting the \texttt{ST-U} model (SE 0.3, ER 0.1, LP $10^4$, MM on), while panels E and F refers to the main mode found with \texttt{ST+PDT} and \texttt{PDT-U} models (SE 0.8, ER 0.1, LP $10^3$, MM off) (for these last two, low-resolution, LR, \XPSI settings have been adopted). 
    }
    }
    \label{fig:config}
    \end{figure*}
}

{
    \begin{figure*}[t!]
    \centering
    \includegraphics[
    width=18cm]{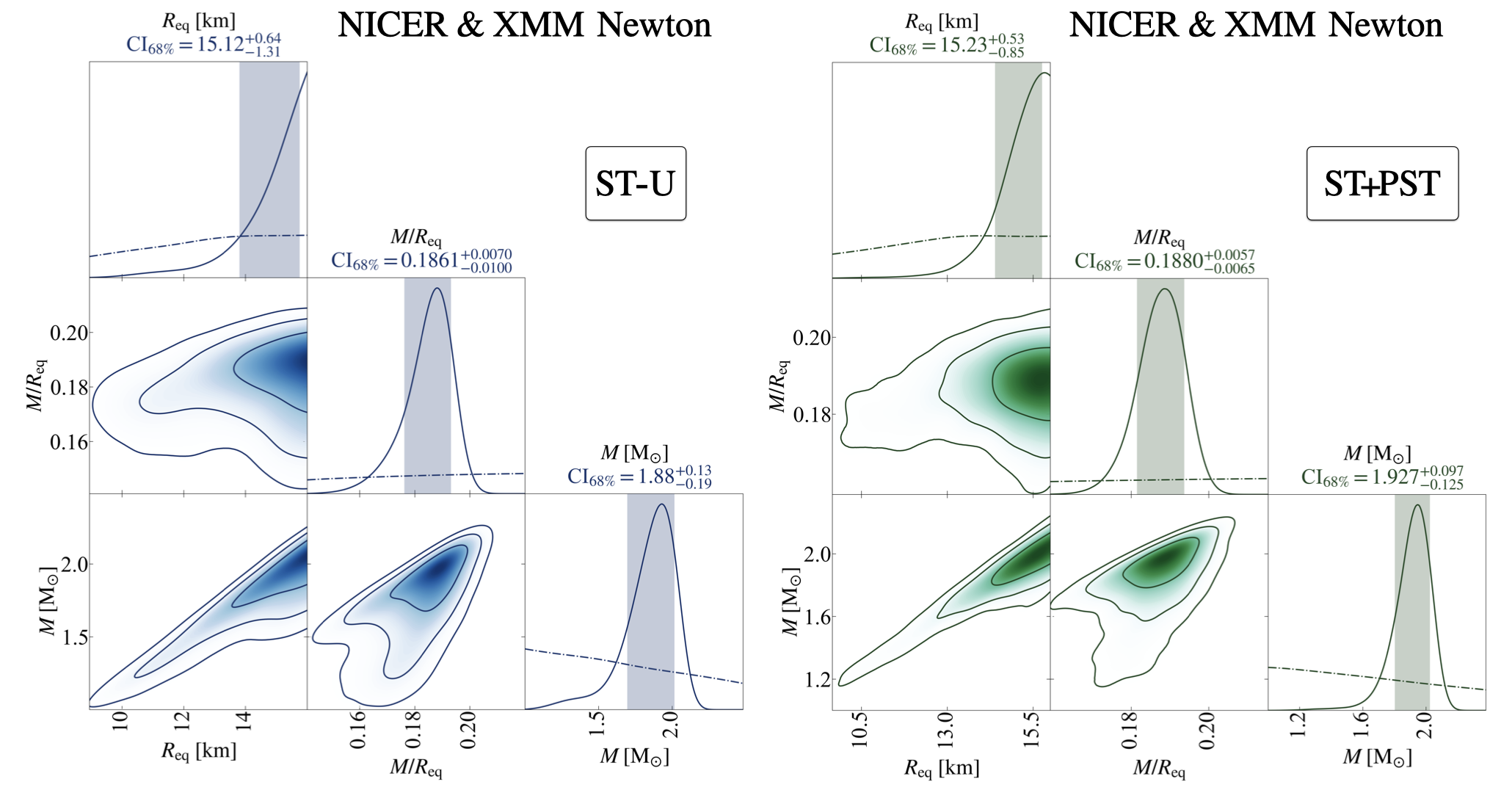}
    \caption{\small{Posterior distributions (smoothed by GetDist KDEs) of radius, compactness, and mass from joint \NICER and \xmm inference runs: left and right corner plots respectively obtained adopting the \texttt{ST-U} and the \texttt{ST+PST} model. With dash lines here we show the prior distributions. In the 2D posteriors we plot 3 curves, defining the 68\%, 95\%, and 99\% credible area. Analyses settings for the depicted runs are displayed in Table \ref{tab:runs} (see \texttt{ST-U} and \texttt{ST+PST} entries with ``yes'' in the \xmm column). 
    See caption of Figure \ref{fig:STU_res} for further details. } 
    }
    \label{fig:STU_STPST_NxX}
    \end{figure*}
}

{
    \begin{figure*}[t!]
    \centering
    \includegraphics[
    width=16cm]{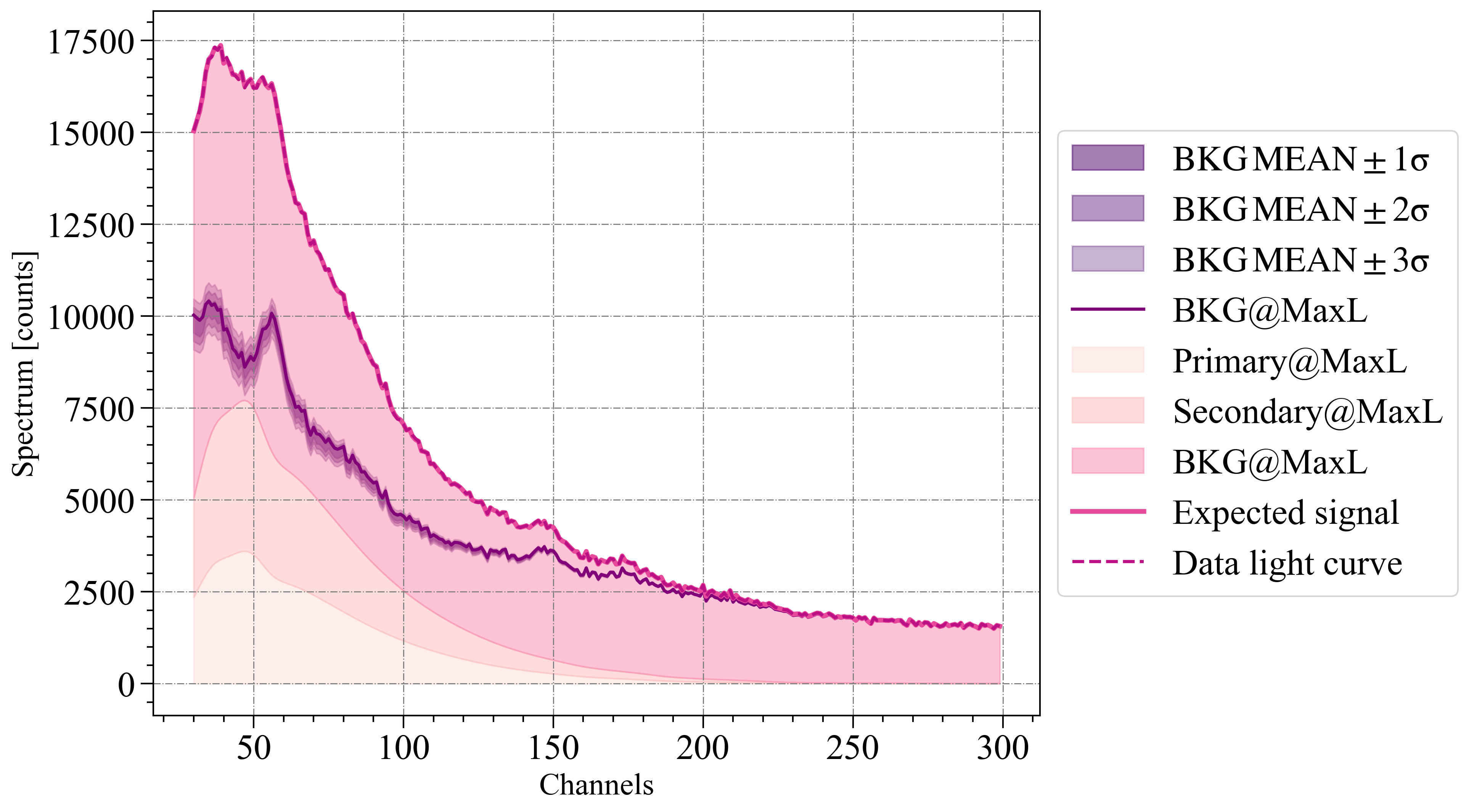}
    \caption{\small{ 
    Spectrum of \NICER data and its inferred composition, according to the maximum likelihood sample of our \XPSI analysis.  
    Data are plotted with a dashed magenta line;
    the total signal expected is shown with a  solid pink line; 
    the background (BKG) that maximises the likelihood for the (background-maginalised) maximum likelihood posterior sample  (MaxL, in the legend) is plotted with a purple solid line. 
    The contribution of the primary hot spot, the secondary hot spot, and the background, again associated with the maximum likelihood sample, are reported on top of each other with pink-shaded regions, increasingly more intensely colored. 
    The purple shades show from darker to lighter color the average background $\pm$ 1, 2 or 3 standard deviations, calculated over 200 randomly selected posterior samples from the equal weight
    \MultiNest output file. 
    The inference runs for this figure concern the joint \NICER and \xmm analysis, performed with the \XPSI \texttt{PDT-U} model. 
    The complete figure set (8 images) is available in the online journal, for the \texttt{ST-U}, \texttt{ST+PST}, \texttt{ST+PDT}, \texttt{PDT-U} models for both \NICER-only analyses (reference runs highlighted in bold in Table \ref{tab:runs}) and joint inferences for \NICER and \xmm data sets (analysis settings reported in Table \ref{tab:runs}, corresponding to ``yes'' entries in the \xmm column).
    } 
    }
    \label{fig:BKG}
    \end{figure*}
}

{
    \begin{figure*}[t!]
    \centering
    \includegraphics[
    width=18cm]{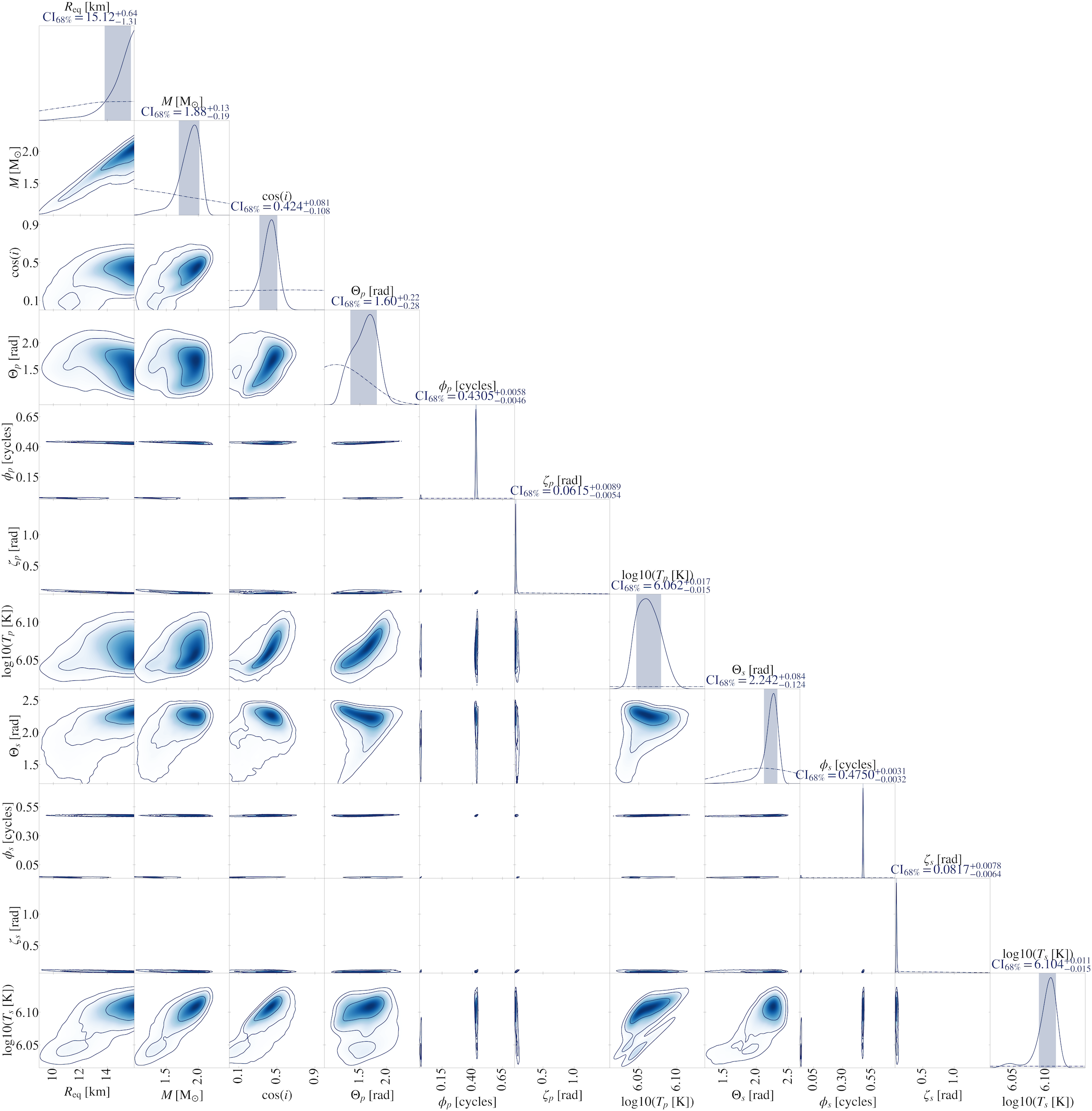}
    \caption{\small{ 
    Corner plot based on the joint analysis of \NICER and \xmm data sets, carried out with the \texttt{ST-U} model. 
    This image includes the posterior distributions (smoothed by GetDist KDEs) for the parameters describing mass, radius, and the hot spot properties inferred for \joo's system. 
    In the marginalized 1D posteriors we show with dash lines our corresponding priors. 
    The complete figure set (8 images) is available in the online journal, for the \texttt{ST-U}, \texttt{ST+PST}, \texttt{ST+PDT}, \texttt{PDT-U} models for both \NICER-only analyses (reference runs highlighted in bold in Table \ref{tab:runs}) and joint inferences for \NICER and \xmm data sets (analysis settings reported in Table \ref{tab:runs}, corresponding to ``yes'' entries in the \xmm column). 
    Tables \ref{tab:params} and \ref{tab:add_params} show the meaning of the adopted parameter labels. 
    The multi-modal structure of the posterior, described in Section \ref{subsubsec:STU_NxXMM}, is here clearly visible. In particular the bi-modality in the 2D posteriors including the radius, shows the main mode (configuration corresponding to its maximum likelihood sample shown in Panel C of Figure \ref{fig:config}) and a secondary one (configuration corresponding to its maximum likelihood sample shown in Panel D of Figure \ref{fig:config}). 
    For more general details about the plots in this figure set, see Figure \ref{fig:STU_res}.
    } 
    }
    \label{fig:params}
    \end{figure*}
}

\subsubsection{Analysis of XMM-Newton Data Only}
The \xmm data of \joo was first analyzed in  \citet{BogdanovGrindlay2009} in an attempt to extract information about the \ac{NS} radius. In this analysis, which employed a frequentist approach, a fixed mass of $M=1.4\,\mathrm{M_\odot}$ was assumed and just $R$ was allowed to vary. Only two circular hot spots were considered with a configuration equivalent to the \texttt{ST-U} model.  In addition, the pulse profile obtained from the \xmm EPIC pn 
(used in \citealt{BogdanovGrindlay2009}, but not in this work, see Section \ref{subsec:dataset_XMM})
instrument was fit in only two broad energy bands (0.3–-0.7\,keV and 0.7–-2\,keV). Due to the limited photon statistics of the \xmm data, this analysis resulted in only an upper limit on the neutron star radius of  $R > 10.7$\,km (95\% confidence) and very broad constraints on the spot locations and geometry, which are generally consistent with configurations C and D in Figure~\ref{fig:config}. 

Using the pipeline and procedures outlined in this paper, analysis of only the \xmm data yields
only very weakly constrained hot spot properties and 68\% credible intervals of $R_{\mathrm{eq}} = 11.3\pm 2.5$\,km and mass $M = 1.6^{+0.4}_{-0.3}\, \mathrm{M_\odot}$. 
These are broader than the values quoted in \citet{BogdanovGrindlay2009}, likely because some of the assumptions have been relaxed and no timing information has been used.  They are also much wider  than the posterior distributions derived from \NICER data, as expected due to the much smaller \xmm exposure time, effective area, time and energy resolution.

\subsection{\texttt{ST+PST}}
\subsubsection{Reproducing R19 Results and Analysing the Impact of Settings}
\label{subsubsec:STPST-R19}
For \texttt{ST+PST}, \citetalias{Riley2019} reported credible intervals for mass and radius respectively of $1.34^{+0.15}_{-0.16}\, \mathrm{M_\odot}$ and $12.71^{+1.14}_{-1.19}$\,km. 
These results were obtained from a run adopting the following \MultiNest settings: SE 0.3 (0.8), ET 0.1, LP $10^3$, MM off, where the sampling efficiency was increased to 0.8 when resuming the analysis. 
In Figure \ref{fig:STPST_res_sett}, we report posterior distributions derived analysing B19v006 with the \texttt{ST+PST} model in the new inference framework (see Section \ref{sec:method} for more details). 
The green lines on the left corner plot show results for the runs that match as much as possible the analysis settings adopted in \citetalias{Riley2019}. 
We notice that there is some scatter among these posterior distributions. In particular the major outlier only finds the secondary mode of the posterior surface (see Sections \ref{subsubsec:STPSTexplore} and \ref{sec:discussion} for context and discussion over its multi-modality); this resembles the mode found for \texttt{ST-U} and is therefore characterized by smaller radii and masses. 
The other two runs still present some scatter, suggesting that these \MultiNest settings are now (given the various changes to channel choices, instrument response etc. compared to \citetalias{Riley2019}) inadequate to explore the model parameter space. 
Compared to the credible intervals presented in \citetalias{Riley2019}, the new posteriors show slightly higher values (likely due to the difference in channels, as the same trend was found in preliminary analyses and attributed for the \texttt{ST-U} model to the removal of data in channels 25 to 30) and similar uncertainties. 
The only HR run with SE 0.3-only is the inference run which identifies as its main mode, the secondary (\texttt{ST-U}-like) mode (1D and 2D posteriors for mass, radius and compactness are represented in Figure \ref{fig:STPST_res_sett} by the green outlier); the computational cost of this analysis was 3-4 times the cost of the two other HR runs.

The same corner plot also shows the effect of lowering the resolution of \XPSI settings; the green curves show the only posteriors obtained with high resolution for models more complex than \texttt{ST-U}. 
The yellow lines represent distributions, derived with low resolution but with the other settings directly comparable to those obtained with high resolution. 
Note that the two outliers are produced with the same fixed \MultiNest and Python seeds. 
Excluding the outlier (which is again connected to the identification of the secondary mode)
 the low-resolution runs seem to converge to slightly more stable solutions located at slightly smaller radii and masses and are characterised by marginally larger posteriors, when the mode favoured by the evidence\footnote{ I.e. the mode identified with more adequate \MultiNest settings. That mode is indeed clearly significantly contributing to the estimation of the evidence, as when identified the evidence significantly increases.} is identified (see Section \ref{subsubsec:STPSTexplore}). 
The behaviour is different when only the secondary mode is explored by the sampler; comparing the two outliers in the 1D radius plot, indeed, the high-resolution run produces much narrower posteriors, peaking at significantly lower values. 
We use low resolution also for our new reference \texttt{ST+PST} run, obtained increasing the number of live points to $10^4$ and enabling the mode-separation option of \MultiNest. 
We report the corresponding posterior with a black dashed line in the same left corner plot of Figure \ref{fig:STPST_res_sett}. 
Compared to the results found with the other low-resolution runs (where the number of \MultiNest live points was $10^3$), these distributions slightly widen and move toward lower values of mass and radius. 
Similar widening of the posteriors, with increasing number of \MultiNest live points, was also found in \citet{Riley2021} for \joh and could indicate a broader true posterior distribution. 
The 68\% credible intervals of mass and radius derived from  our new reference inference are: $1.37\pm 0.17\,\mathrm{M_\odot}$ and $13.11\pm 1.30$\, km (as reported on top of the 1D posteriors in the left corner plot of Figure \ref{fig:STPST_res_sett}), largely in agreement with the findings of \citetalias{Riley2019}. 
The hot spot configuration corresponding to the maximum likelihood sample of this analysis is reported in panel B of Figure \ref{fig:config}. 
Our current tests do not guarantee that these analysis settings are sufficient to adequately explore the model parameter space.

In the right corner plot of Figure \ref{fig:STPST_res_sett}, we show posterior distributions derived from each of the low-resolution runs, reported with yellow lines in the left panel. 
We also report in ocher the posterior identified by \citet{Riley2019} for the \texttt{ST+PST} model. 
The outlier, depicted in the right corner plot with blue lines, is found by a replica of an inference run with our default \MultiNest settings. Hence the different results can be attributed solely to the randomness present in the sampling process. 
The occurrence of such outlier, however, suggests that the number of live points chosen for these inferences is not  sufficient to effectively explore the model parameter space. 
The black dashed curves represent the results obtained when introducing the elsewhere temperature to the \texttt{ST+PST} model and are presented in more detail in Section \ref{subsubsec:STPSTexplore}.

The other three inference runs 
show posterior distributions almost perfectly overlapping, despite the differences in \MultiNest settings (the yellow lines show the results obtained enabling the mode-separation option) and in the prior (the green curves outline distributions derived from runs adopting the updated \texttt{ST+PST} CoH prior). 
With this plot and the associated inference runs, we demonstrate that, in analysing the \NICER data B19v006, 
introducing the updated \texttt{ST+PST} CoH prior leads to no significant difference in the inferred posterior distributions and evidence estimation. 
We can draw the same conclusion for the effect of adopting the mode-separation \MultiNest option, even though limited to the specific case here considered. 
{
    \begin{figure*}[t!]
    \centering
    \includegraphics[
    width=18cm]{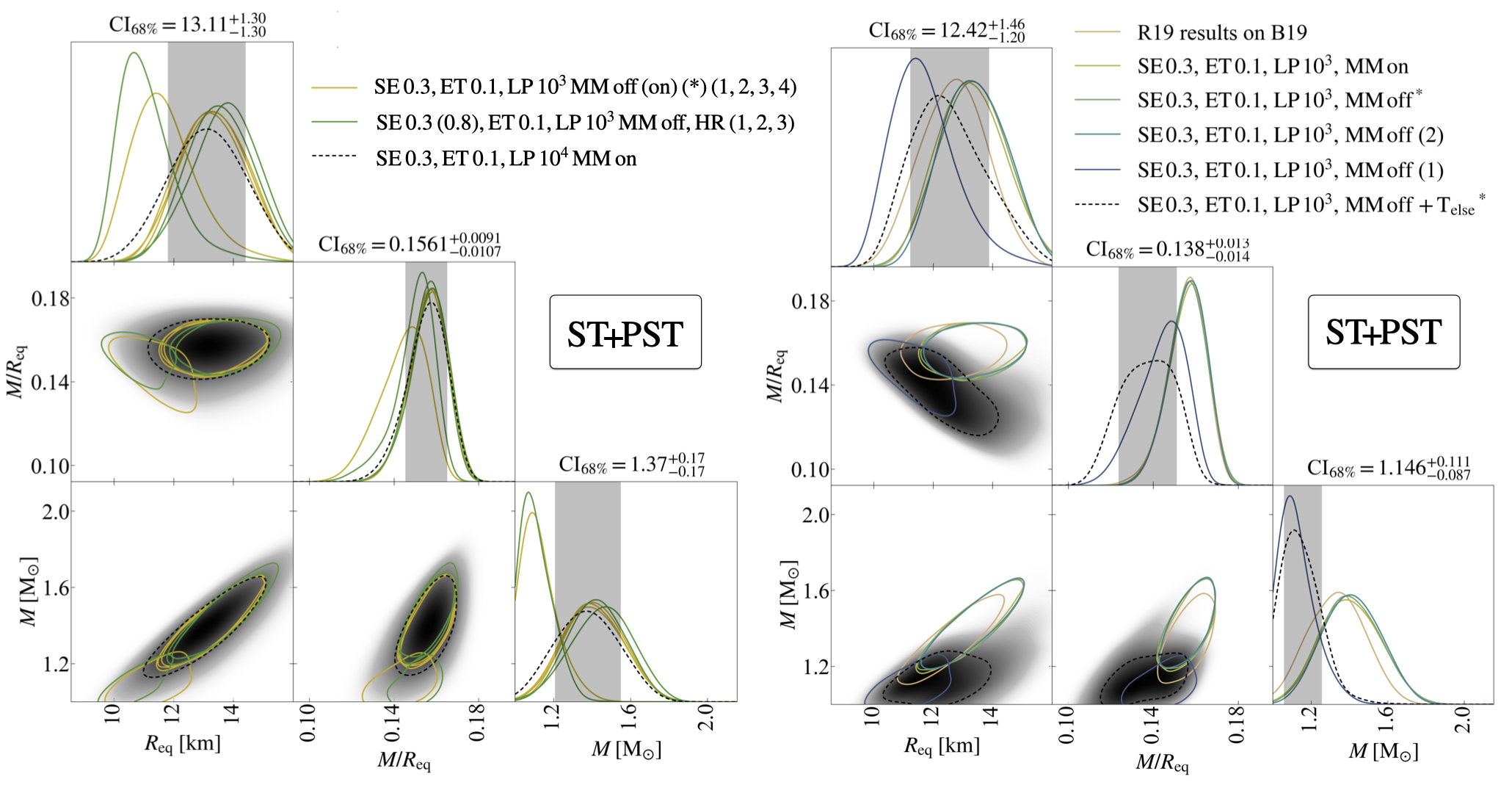}
    \caption{\small{
    \texttt{ST+PST} posterior distributions (smoothed by GetDist KDEs) from 14 runs, for radius, compactness, and mass. This figure highlights the effects of adopting different \MultiNest settings. 
    The corner plot on the left shows the effect of adopting different (low or high) \XPSI resolution settings. 
    The green curves represent the posterior distributions inferred from the analyses that adopt high resolution (HR in the legend); the yellow ones are derived from low \XPSI resolution setting (our default for models more complex than \texttt{ST-U}). These latter have slightly different \MultiNest settings, as marked with settings in parenthesis, but show very similar behaviours and are individually plotted on the right panel. 
    The two outliers (one at HR, green, and one at LR, yellow) have been obtained fixing \MultiNest and Python random seeds to the same values.
    With black dash lines, we also report the posterior distributions obtained with the SE 0.3, ET 0.1, LP $10^4$, MM on run; numbers on top of the 1D posteriors, as well as shaded areas refer to this inference. 
    On the right panel, instead, we show new results for 5 \texttt{ST+PST} inference runs, with low \XPSI resolution, adopting our default \MultiNest settings, the only exception being the second  run from the top of the list, for which we enabled the mode-separation. Similar \MultiNest settings have also been adopted in \citetalias{Riley2019} (although higher \XPSI settings were in that case used). The \texttt{ST+PST} posteriors found in \citetalias{Riley2019} are here shown in brown for comparison. 
    This plot tests: the variability due to the randomness of the sampling process, given the specific analysis settings; the effect of updating the model prior, to allow the primary hot spot to overlap with the masked part of the secondary (CoH, runs marked in the legend with $^*$); and the effect of introducing emission from all the \ac{MSP} surface (see run including $\mathrm{T_{else}}$ in the legend). The black dashed lines represent the posterior distributions obtained when allowing the entire surface of the \ac{MSP} to emit. Shaded areas and numbers on top of the 1D posterior distribution refer to the credible regions and intervals of this inference. 
    See caption of Figure \ref{fig:STU_res} for further details. \\
    $^*$: runs adopting the updated CoH \texttt{ST+PST} prior, as explained in Section 2.3.4 of \citet{Vinciguerra2023a}. 
    }}
    \label{fig:STPST_res_sett}
    \end{figure*}
}

\subsubsection{Model Exploration}
\label{subsubsec:STPSTexplore}
Our mode-separation inference runs highlight the multi-modal structure present in the posterior surface of this specific model, given the chosen B19v006 data set (although we expect all \joo\ \NICER data sets to exhibit similar behaviour).
As briefly mentioned in the previous section, there are two prominent modes in the posterior, which differ in local log-evidence by a factor of $\sim 5.5$. 
The maximum likelihood samples belonging to the two modes actually differ by a much larger value ($>$10 in log-likelihood), from which we can deduce that the prior space supporting the secondary mode is considerably larger than that supporting the main mode. 
This also explains why a relatively low number of live points can result in the identification of the secondary mode as the main one. 
An illustrative hot spot geometry associated with the main mode is rendered, taking as example the maximum likelihood sample, in panel B of Figure \ref{fig:config}. 
The secondary mode resembles, also in the inferred background, the main mode identified with the \texttt{ST-U} model and reported in panel A of the same Figure. 
The omitting component is always associated with the smaller, closer-to-the-equator, hot spot (labeled as primary in panel A of Figure \ref{fig:config}). 
The location and size of the masking element can vary significantly within the identified mode. 
Analysing the SE 0.3, ET 0.1, LP $10^3$, MM off, LR run that missed the main mode, but found this secondary one (whose mass, compactness and radius posterior are represented with blue lines in the right corner plot of Figure \ref{fig:STPST_res_sett}), we notice that, while the maximum likelihood sample of this run depicts the primary as a ring, the maximum posterior sample represents it as an almost circular hot spot, where the omitting component barely touches the emitting one.
This spread in the \texttt{PST} hot spot characteristics suggests, as found in \citetalias{Vinciguerra2023a}, that we are 
only very weakly sensitive to the small details of the hot spot shapes. 
In this case, it seems that we are mostly sensitive to the location and the overall emitting area of the \texttt{PST} hot spot: there is indeed a clear correlation between the size of the masking spherical cap and its distance from the center of the emitting one.  

Both primary and secondary modes deliver a very similar inferred background (see Figure 11 of \citetalias{Vinciguerra2023a}, for a comparison between the background associated with \texttt{ST+PST} and \texttt{ST-U} main solutions) and compactness values, despite the differences in the inferred mass and radius (comparable compactness values are also deduced with our reference \texttt{ST-U} analysis). 
In other aspects these two solutions are quite different from each other. 
In particular the difference in the inferred observer inclination, estimated by the two modes, yields changes in the hot spot location and sizes, such that they are still observable at the right phases (a more detailed discussion on the effect of different inclinations can be found in Section 5.2.1 of \citetalias{Vinciguerra2023a}). 
These considerations highlight the presence of significant, and sometimes non-trivial, degeneracies in our model parameter spaces.

In the right plot of Figure \ref{fig:STPST_res_sett}, we show the effect of allowing the rest of the surface of the star (the part not covered by the two hot spots) to emit black-body radiation characterised by a finite, homogeneous temperature (\texttt{ST+PST}$+T_{\mathrm{else}}$). 
This inference also uses the updated CoH prior, however, as mentioned above, we do not expect this choice to have a significant impact on the final results. The credible intervals on top of the 1D plots refer to the results of this inference. The configuration found for this model, with these \MultiNest and \XPSI settings, resembles that obtained with the \texttt{ST-U} model and represented, for the maximum likelihood sample of that inference run, in panel A of Figure \ref{fig:config}. The mask is again associated with the hot spot being closer to the equator (labeled as primary in panel A) and changes its shape to rings and crescents (both maximum likelihood and posterior exhibit ring-like shapes for this secondary hot spot).  

Allowing all of the surface of the \ac{MSP} to emit introduces more uncertainty on the background and increases uncertainty on the inferred compactness. 
We find an almost flat posterior support for elsewhere temperatures below $\sim 10^{5.6}\,$K, 
likely due to the negligible contribution of black-body radiation of such low temperatures within \NICER band (for comparison the temperature of the two emitting components are 
$T_{\mathrm{p}}\sim T_{\mathrm{s}}\sim 10^{6.1}\,$K). 
This additional component can explain part of the unpulsed 
emission that is otherwise mostly attributed to the background and to higher compactness values \citep{Riley2021,Salmi2022}. 
This degeneracy between these three modeled components ($T_{\mathrm{else}}$, background and compactness), and between $T_{\mathrm{else}}$ and the hydrogen column density (whose 68\% credible interval has indeed increased and widen, compared to the runs without $T_{\mathrm{else}}$ that identified as main mode the \texttt{ST-U}-like mode), generates the aforementioned uncertainties, which are e.g. visible comparing the blue with the black dashed lines in the right corner plot of Figure \ref{fig:STPST_res_sett}. 
The slight cut in high compactness values, in combination with the increased uncertainties, yields a posterior that peaks at larger radii (compared to the other runs which identified the \texttt{ST-U} like mode as the main mode). 
Overall, the most prominent change generated by the elsewhere temperature is the broadening of the posteriors (in addition to doubling the computational time).

\subsubsection{Joint Analysis of NICER and XMM-Newton Data}
We report the mass, compactness, and radius posterior distributions obtained adopting the \texttt{ST+PST} model in a joint \NICER and \xmm inference analysis in the right corner plot of Figure \ref{fig:STU_STPST_NxX}. 
These results were obtained by a SE 0.8, ET 0.1, LP $10^3$, MM off, LR run in terms of \XPSI settings. 

As is immediately clear, comparing the two corner plots of Figure \ref{fig:STU_STPST_NxX}, this \texttt{ST+PST} inference run leads to radius, compactness, and mass posterior distributions very similar to those derived with the \texttt{ST-U} model. 
The radius posterior is again truncated close to its peak by our prior cut at 16\,km. This impacts the inferred mass distribution which appears to be then driven by constraints on the compactness. 
The similarities are not confined to these parameters, indeed the inferred hot spot geometries are also almost identical 
(as seen in the online set of Figure \ref{fig:params}),
 and therefore resemble the configuration depicted in Panel C of Figure \ref{fig:config}.  
As shown in Panel C, both hot spots have small sizes and similar temperatures. 
Small sizes lead to (type I, as explained in \citetalias{Riley2019}) degeneracies, when we assume complex hot spot geometries (i.e. when the hot spot is described by two components, as it is the case for  our \texttt{PST} hot spot). 
In practice, the inferred small sizes of the hot spots imply a small sensitivity to their shape. Therefore this yields weak constraints on the omitting component, which, given the similarities between the two hot spots, can be assigned to either of them. 
Since in this case we define the \texttt{PST} hot spot as secondary, this ambiguity gives rise once again to visible bimodality in the posterior distributions of the parameters describing the properties of the two hot spots (see again the online set of Figure \ref{fig:params}). 

As described in Section \ref{subsubsec:STU_NxXMM}, to jointly explain both the phase-resolved \NICER data and the \xmm data, it is necessary to increase the contribution of \joo to the total counts. An increase of the unpulsed emission from the \ac{MSP} can be produced by higher compactness values, as inferred here. 

The similarities in the mass and radius posteriors with those obtained with the \texttt{ST-U} model, extend to the tails of these distributions. 
They thus hint at the presence of a secondary mode, again characterised by lower radius and mass values (respectively $\sim 11$\,km, and $\sim1.4\,\mathrm{M_\odot}$). 
The samples belonging to these tails display again (see panel D of Figure \ref{fig:config},  for a qualitative representation) higher inclination, with both hot spots increasing slightly in size and migrating toward the equator, compared to the main mode. 
Also in this case, we find configurations in which the location of the \texttt{ST} and \texttt{PST} hot spots alternate. 
Depending on the size of the omitting component (which is not well constrained), the emitting region masked by it may slightly increase its size, to compensate for the otherwise decreased emission. This spread over the possible properties of the omitting region highlights again our weak sensitivity to the details of hot spot shapes.

\subsection{\texttt{ST+PDT}}
\subsubsection{Model Exploration}
In Figure \ref{fig:STPDT}, we show the posterior distributions inferred for mass, radius, and compactness, adopting the \texttt{ST+PDT} model (a more complex model than the \texttt{ST+PST} model used for the headline result in \citetalias{Riley2019}, that was not explored in that paper). 
The left corner plot refers to the results derived with the SE 0.8, ET 0.1, LP $10^4$, MM on, LR run, analysing only the B19v006 \NICER data set. 
The inferred 68\% credible intervals for mass and radius are  $1.20^{+0.14}_{-0.11}\,\mathrm{M_\odot}$ and $11.16^{+0.90}_{-0.80}$\,km. 
This model therefore identifies quite a narrow peak in radius. 
The associated hot spot configuration and temperatures resemble the one reported in panel E of Figure \ref{fig:config}. 
The bulk of the inferred posteriors describe hot spot patterns, mass, and radii similar to
those identified as secondary mode in the joint analysis of \NICER and \xmm data sets with the \texttt{ST-U} model \footnote{Note that the \texttt{ST+PST} model shows also similar posterior samples in its tails, suggesting the presence of a secondary mode.}.  
We find considerable differences, however, in the two temperatures associated with the \texttt{PDT} hot spot, compared to the temperatures associated with \texttt{ST} emitting regions derived with the combined \NICER and \xmm analyses. 
Similar emission to the $\log_{10}(T_{s}/\mathrm{K})\sim 6$ \texttt{ST} hot spot in the \texttt{ST-U} model, is often modeled here, as a \texttt{PDT} hot spot, with one component slightly colder ($\sim 10^{5.9-6.0}\,$K) and larger and one tiny but very hot ($\sim 10^{6.3-6.4}\,$K). 
The peak of the posterior identifies the ceding region with the small hot spherical cap, however, there is some posterior mass that also associates it with the cold and more extended component. 
This bimodality is also visible, in the 95\% credible area contours, in the 2D posterior distributions of the parameters describing the properties of the secondary hot spot (see Figure set \ref{fig:params} in the online journal). 
Unlike the results derived from \texttt{ST-U} and \texttt{ST+PST} models in joint \NICER and \xmm analyses, the posterior distributions identified with \texttt{ST+PDT} show no bimodality generated by label ambiguity. 
This means that there is a significant preference for the location of the \texttt{PDT} hot spot; i.e. the data are better represented if one specific emitting region (the first visible, in our representation of the data, see Figure \ref{fig:data}) is described by two components with different temperatures. 

With the SE 0.8, ET 0.1, LP $10^4$, MM on, LR run, we also find a secondary mode. 
This mode shows a maximum likelihood that differs from that of the main mode by $\sim 4$ in log. 
The differences in local evidences are slightly larger, $\sim 6$ in log. 
The secondary mode identifies configurations resembling (with some variation) the main mode identified with the \texttt{ST-U} model when analysing \NICER-only data. 
In this case the dual temperature hot spot may be associated with either of the two emitting regions. The temperature of the superseding component of the \texttt{PDT} hot spot remains of the same order as found with the \texttt{ST-U} model $\log_{10}(T_{s}/\mathrm{K})\sim 6.1$ and is similar to the temperature associated with the \texttt{ST} hot spot. 
The ceding component is considerably colder $\log_{10}(T_{c,s}/\mathrm{K})\sim 5.2-5.6$. 
The resemblance to the \texttt{ST-U} \NICER-only case also includes the mass and radius posteriors, which cluster around $\sim 1.1\,\mathrm{M_\odot}$ and $\sim 10.5\,\mathrm{km}$, slightly lower than the values for the \texttt{ST+PDT} main mode.

In Section \ref{subsec:caveats}, 
we will elaborate on the adequacy of the coverage of the model parameter space during the inference analyses presented in this Section. 
{
    \begin{figure*}[t!]
    \centering
    \includegraphics[
    width=18cm]{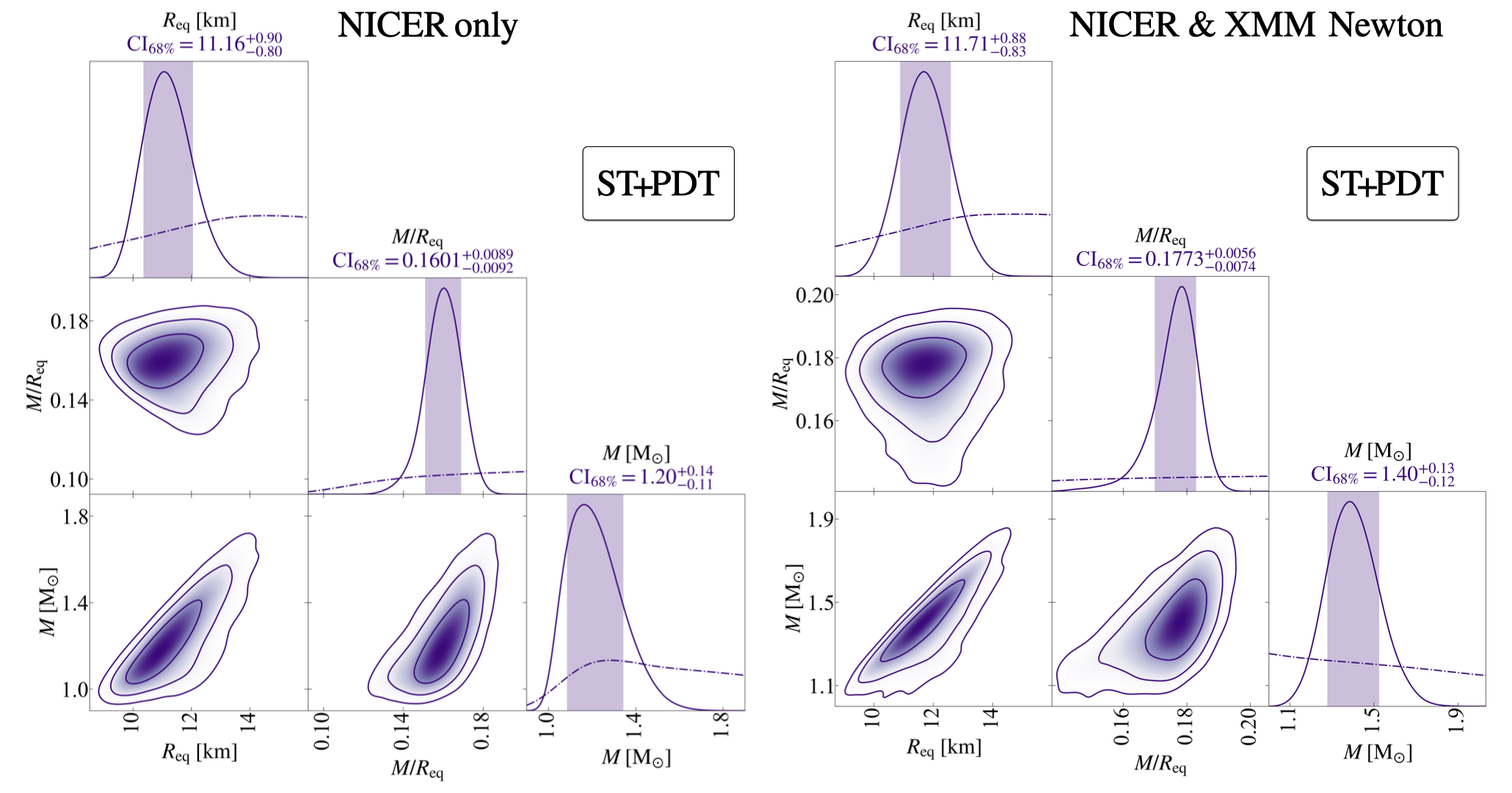}
    \caption{\small{
    \texttt{ST+PDT} posterior distributions (smoothed by GetDist KDEs) of radius, compactness and mass for two inference runs. 
    We show results obtained with the analysis of \NICER-only data (left), and derived by the joint analysis of \NICER and \xmm data (right). 
    For more details about the plot see Figure \ref{fig:STU_STPST_NxX}. 
    } 
    }
    \label{fig:STPDT}
    \end{figure*}
}
\subsubsection{Joint Analysis of NICER and XMM-Newton Data}
The effect of introducing \xmm data into the inference process, when adopting the \texttt{ST+PDT} model, is shown in the right corner plot of Figure \ref{fig:STPDT}. 
The reported posterior distributions were derived from a SE 0.8, ET 0.1, LP $10^3$, MM off, LR run. 
In panel E of Figure \ref{fig:config}, we show the hot spot geometry of the identified maximum likelihood sample. 

For the first time, coherently adding the \xmm data set in the inference process does not yield solutions that are visibly different from those obtained analysing only \NICER data.  As with other models, the introduction of the \xmm data constrains the background to slightly lower values than otherwise inferred. 
In this case, we also see an increase in the compactness, although the difference is significantly smaller when compared to the other models. 
The properties of the pulsed emission are then recovered by increasing both mass (whose 68\% credible interval is now $1.40^{+0.13}_{-0.12}\,\mathrm{M_\odot}$) and radius (whose 68\% credible interval is now  $11.71^{+0.88}_{-0.83}$\,km). 
Indeed for both of these parameters the posterior mass at higher values has considerably increased, while the posterior volume for the lowest values has significantly reduced, compared to analysis of the \NICER-only data. 

Looking at the posterior distributions of the hot spot parameters (see  Figure set \ref{fig:params} in the online journal ), only the configuration assigning very high temperature to the superseding component of the \texttt{PDT} hot spot, has been identified by this joint \NICER and \xmm analysis (instead of the predominant alternative found with \NICER-only data, where the tiny hot spherical cap was the ceding component and hence partially hidden by a cooler and larger component). 
There is also a significant anti-correlation between the sizes of the hot spots (particularly of the \texttt{ST} and the superseding component of the \texttt{PDT} hot spot) and inferred mass and radius of \joo. 
Most likely, this correlation can be explained by the necessity of a similar emitting area, which requires a lower angular radius of the hot spot if the \ac{MSP} has a larger radius. 
It is not clear if this is a consequence of the introduction of the \xmm data, or if it is caused by the low number of live points adopted to explore the large \texttt{ST+PDT} parameter space. 
Indeed, as also reported in Table \ref{tab:runs}, the \texttt{ST+PDT} joint \NICER and \xmm analysis was performed with a considerably lower number of live points ($10^3$), compared to the corresponding \NICER-only inference. 
Similarly to the \NICER-only inference, the uncertainty over the colatitude of the \texttt{ST} hot spot, allows it to be located on either the same or the opposite hemisphere as the observer. 
In this joint analysis, two distinct peaks are present in the 1D colatitude posterior; however this could again be due to the relatively low number of live points enabled to explore the parameter space. 

\subsection{\texttt{PDT-U}}
\subsubsection{Model Exploration}
\label{subsubsec:PDTUexplore}
We analyse the revised \NICER \joo data set with a SE 0.8, ET 0.1, LP $10^4$, MM on, LR inference run. 
In the left corner plot of Figure \ref{fig:PDTU}, we report the radius, compactness, and mass posterior distributions derived adopting the \XPSI \texttt{PDT-U} model. 
We find relatively wide posteriors and hence large 68\% credible intervals; the estimated mass and radius are respectively $1.41^{+0.20}_{-0.19}\,\mathrm{M_\odot}$ and $13.12^{+1.35}_{-1.21}$\,km. Uncertainties and values resemble those inferred with the \texttt{ST+PST} model, when \xmm data are not included. 

From the posterior distributions of the parameters describing the geometry and hot spot properties of the system (see corresponding corner plot, Figure set \ref{fig:params} in the online journal), we note the presence of three distinct modes. 
These were also identified through the mode-separation \MultiNest option. 
The main mode is characterised by a hot spot configuration similar to that reported in Panel F of Figure \ref{fig:config}. 
The solutions with the highest likelihood values from this inference run belong to this mode and have relatively low mass ($\sim 1-1.3\,\mathrm{M_\odot}$) and quite high radius ($\gtrsim 15\,$km), populating most of the lower tail of the observed 1D compactness posterior distribution. However, this mode extends to  and dominates also other parts of the parameter space, spanning a relatively wide range in compactness. 
This spread includes considerably smaller radii and significantly higher masses. 

Within this mode, both hot spots are characterised by a cold and large ceding component and a hot, small superseding component. 
The secondary hot spot (as for all the \XPSI models ending with \texttt{-U}, this refers to the hot spot with higher colatitude) typically coincides with the surface area generating the first emission peak recorded in the \NICER data, according to our representation (see Figure \ref{fig:data}). 
The secondary hot spot is characterised by both the hottest ceding (with average and standard deviation $\log_{10}(T_{\mathrm{c,s}}/\mathrm{K})\sim 5.86\pm 0.04$ vs $\log_{10}(T_{\mathrm{c,p}}/\mathrm{K})\sim 5.7\pm 0.1$ of the primary) and superseding component (with average and standard deviation  $\log_{10}(T_{\mathrm{s}}/\mathrm{K})\sim 6.23\pm 0.03$ vs $\log_{10}(T_{\mathrm{p}}/\mathrm{K})\sim 6.08\pm 0.03$ of the primary). 
In contrast to what was found for the models analyzed in \citetalias{Riley2019}, both hot spots are typically located in the hemisphere directly facing the observer (as shown panel F of Figure \ref{fig:config}). 
There is however some spread on their exact location and in the inferred inclination (average and standard deviation of $\cos(i)\sim 0.3\pm0.2$); in particular if the observer inclination compared to the rotation axis approaches $\sim 0\,\deg$, the hot spots tend to move towards the equator, to maintain the observed pulsed emission. 
Slight variations within this main mode also dominate the tails seen in the 2D posterior distributions reported in the left corner plot of Figure \ref{fig:PDTU}.  

Given the relatively similar colatitude values inferred for the two hot spots and their uncertainties, it is natural to expect a bimodality due to the ambiguity of the primary and secondary labels. 
The posterior distributions of the parameters describing the hot spot properties clearly show this feature.
The secondary mode shows again, for both hot spots, a colder and larger ceding component and a hot, very small superseding component. 
The temperature ranges characterising the hot spot components are similar to the main mode, but the warmer hot spot is now labeled as the primary. Still, the warmer hot spot is responsible for the first emission peak, given the definition of phase zero in the data set that we use. 
However this secondary mode, formally identified also thanks to the mode-separation \MultiNest option, additionally displays slightly different overall features. 
Both hot spots are now more typically located in the hemisphere opposite to the observer, whose inclination compared to the rotation axis is now favoured at 
slightly higher angles 
(with average and standard deviation of $\cos(i)\sim 0.2\pm0.1$). 

The average and standard deviation of mass and radius posterior samples associated with this secondary mode are $1.46\pm0.16\,\mathrm{M_\odot}$ and $12.6\pm0.9$\,km, and therefore populate the main posterior peak of the compactness. 
Although these values do not significantly differ compared to the one inferred from the main mode, in this case these are also representative of the highest likelihood posterior samples.  

The maximum likelihood associated with samples belonging to the secondary and the main mode respectively differ by only 2 in units of log-likelihood, while the difference in local evidence amounts to $\sim3$ units in log. 
The background associated with these main two modes displays some variability, but it is considerably smaller than that inferred for the \texttt{ST-U} and \texttt{ST+PST} solutions. 

Our inference run also identified a third mode, which is visible in some of the 2D posterior distributions corresponding to the parameters describing the hot spot properties. 
Its contribution is however marginal, as confirmed by the best likelihood values associated with this mode. 
They are worse by about $\sim 9$ units in log compared to those associated with the main mode (the difference in local evidence is about $\sim8$ units in log). 
This third mode is the \texttt{PDT-U} representation of the main \texttt{ST-U} mode that was found when analysing only \NICER data. Similarly to that case, the inferred posterior samples cluster around mass and radius values of $1.15\pm0.10\,\mathrm{M_\odot}$ and $10.7\pm0.9$\,km, expressed as average values with standard deviation uncertainties. 
As expected given this resemblance, the background associated with samples belonging to this mode is considerably higher than that inferred from the other two modes. 

The overall posterior distributions of this third mode feature hot spot configurations similar to the main \texttt{ST-U} solutions and their corresponding solution within the \texttt{ST+PST} model parameter space (in that context, describing the secondary mode). 
As in the latter case, the primary, and now also the secondary, exhibits some variation in terms of the relation between the two components, in particular in the characterisation of the cold ceding component (average and standard deviation of $ \log_{10}(T_{c,p/s}/\mathrm{K})\sim 5.5\pm0.3$ for both hot spots). 
Focusing instead on the maximum likelihood solutions belonging to this mode, the complexity introduced to describe the hot spot leads to the presence of a very small ($\zeta_{s}\sim 0.04$\,rad) and hot ($\log_{10}(T_{c,s}/\mathrm{K})\sim 6.6$) ceding component, almost completely masked by the superseding one.

{
    \begin{figure*}[t!]
    \centering
    \includegraphics[
    width=18cm]{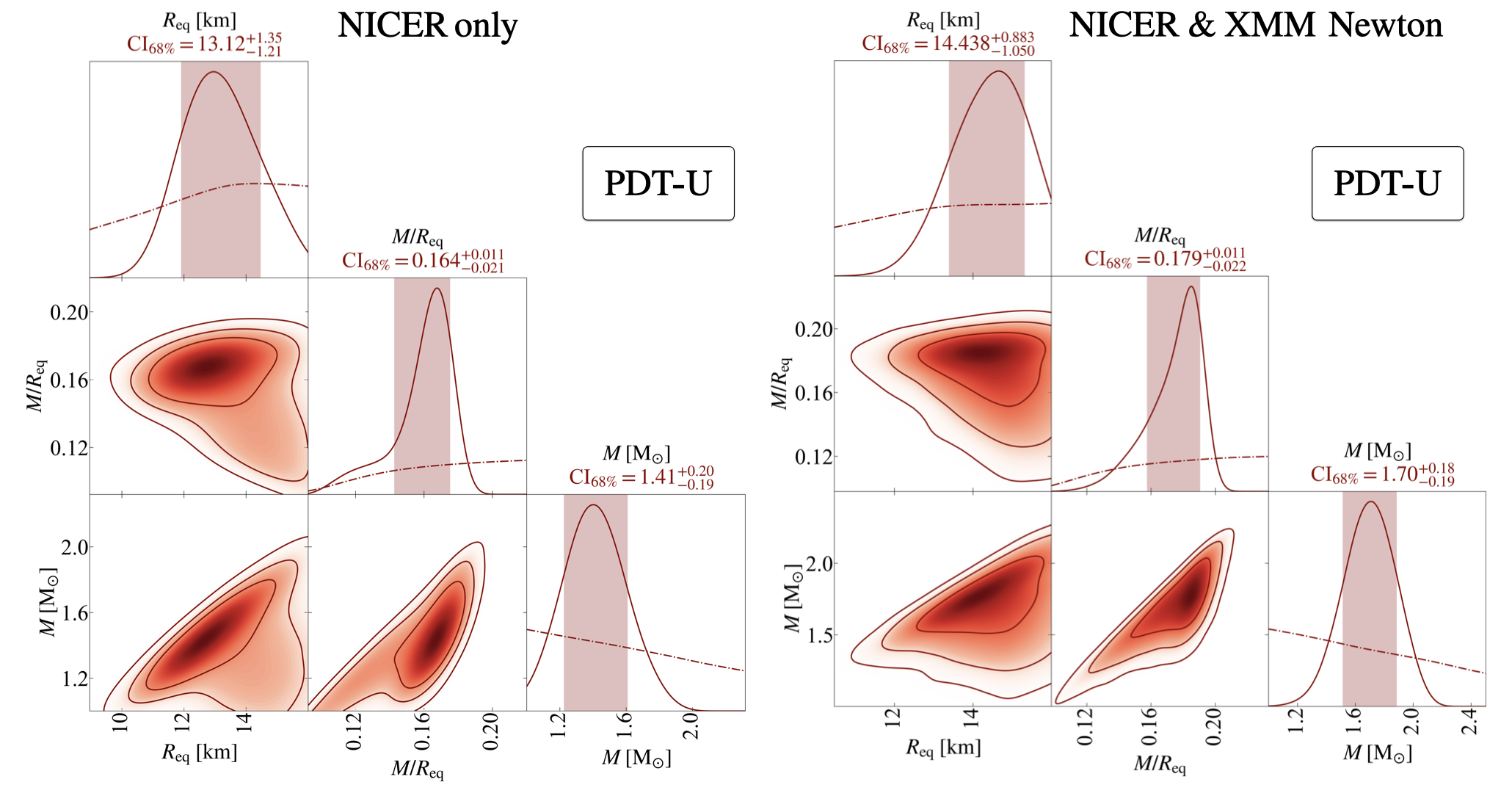}
    \caption{\small{\texttt{PDT-U} posterior distributions (smoothed by GetDist KDEs) of radius, compactness, and mass for two inference runs. 
    We show results obtained with analysis of \NICER-only data (left) and derived by the joint analysis of \NICER and \xmm data (right). For more details about the plot see Figure \ref{fig:STU_STPST_NxX}.
    } 
    }
    \label{fig:PDTU}
    \end{figure*}
}

\subsubsection{Joint Analysis of NICER and XMM-Newton Data}
In the right corner plot of Figure \ref{fig:PDTU}, we show the posterior distributions of radius, compactness, and mass inferred by the joint analysis of \NICER and \xmm data sets. 
These results were obtained through the SE 0.8, ET 0.1, LP $10^3$, MM off, LR run. 

Again the main effect of introducing \xmm data in the analysis consists in increasing the fraction of photons recorded by \NICER, assigned to the emission from \joo. 
Consequently, posterior samples that are associated with high background are here considerably downplayed, and the compactness distribution shifts towards slightly higher values. 
In this case, however, the identified solutions that can reproduce both data sets are qualitatively similar to those obtained analyzing only \NICER data; more specifically belonging to the main and secondary modes of the NICER-only run. 
The same two modes originate from the bimodality structure clearly visible in some of the posterior distributions describing the properties of the two hot spots (see Figure set \ref{fig:params} in the online journal). 
Samples associated with the least prominent mode of the \NICER-only inference, on the other hand, are unable to represent the data given the additional information provided by \xmm. 

In Panel F of Figure \ref{fig:config}, we show, as example, the hot spot configuration associated with the maximum likelihood sample of the joint \NICER and \xmm inference run. 
The geometry of the system is qualitatively representative of the main mode of the posterior, which  resembles the main mode inferred with only \NICER data. 
The maximum likelihood samples reach quite high values for the radius, typically $\gtrsim 15$\,km. 
However, now the associated masses are considerably larger, and cluster around $1.2-1.5\,\mathrm{M_\odot}$. 
The secondary mode (whose qualitative characteristics are described in Section \ref{subsubsec:PDTUexplore}) also migrates towards higher masses, once \xmm data are included in the analysis. The maximum likelihood samples belonging to this mode show masses around $\sim (1.7-2.0)\,\mathrm{M_\odot}$; 
the corresponding radius also shifts toward higher values,
in the range of $\sim(13-16)$\,km. 

Given the relatively high values of radius associated with both modes, we decided to inspect the solutions that form the posterior mass for equatorial radii $<12.5$\,km. 
The samples populating this part of the parameter space exhibit characteristics belonging to both modes. 
At least one of the hot spots (usually the primary) is often located in the hemisphere facing the observer, behaviour  associated with the main mode.  
However, the observer inclination with respect to the spin axis is quite large, a characteristic more often associated with the secondary mode. 

Medians and 68\% credible intervals estimated for mass and radius according to the joint analysis of \NICER and \xmm data sets are $1.70^{+0.18}_{-0.19}\,\mathrm{M_\odot}$ and $14.44^{+0.88}_{-1.05}$\,km respectively. 
The constraint coming from \xmm data shifts the compactness, mass, and radius 
values toward slightly higher distribution.

\section{Discussion}
\label{sec:discussion}
In this Section we discuss and contextualize the results presented in Section \ref{sec:results}. 

\subsection{Comparison with R19 Results}
\label{subsec:STU_comparisonR19}
The results outlined in Sections \ref{subsubsec:STU-R19} and \ref{subsubsec:STPST-R19} are in good agreement with the analyses presented in \citetalias{Riley2019}, when using the same \texttt{ST-U} and \texttt{ST+PST} models.  A few small changes arise due to the differences in set up compared to that adopted in \citetalias{Riley2019}:
\begin{itemize}
    \item Instrument response: for this study we adopt the latest general \NICER instrument response (the one used for the analysis of \joh in \citealt{Riley2021, Salmi2022,Salmi2023}); 
    \item Data sets: as presented in Section \ref{sec:dataset}, the data set analyzed here, B19V006, has been derived from the same event list defined in \citet{Bogdanov19a}, but accounting for the new instrument response;
    \item \NICER channel range used: in this work [30:300), in \citetalias{Riley2019} [25:300);
    \item Priors on $\cos(i)$ and colatitudes: here isotropic and thus sampled from a uniform prior in cosine, in \citetalias{Riley2019} uniformly sampled in angle;
    \item Different modeling of instrument uncertainties: as outlined in Section  \ref{subsubsec:HSmodels}, compared to this study, \citetalias{Riley2019} use two additional parameters to model uncertainties in the instrument response, dealing with the Crab calibration \citep{Ludlam2018};
    \item Updates to the \XPSI pipeline: in this analysis we use \XPSI \texttt{v0.7} and \texttt{v1.0}, whereas \citetalias{Riley2019} used \texttt{v0.1}. For full details of changes see the  CHANGELOG file in the \XPSI GitHub repository\footnote{\url{https://github.com/xpsi-group/xpsi}.};
    \item Settings: both in relation to the \XPSI pipeline and \MultiNest (see in particular differences in live points), as discussed in Sections \ref{subsubsec:ParamPrior} and \ref{subsec:multinest}.
\end{itemize}
One particularly relevant change, observed in the results, concerns the widening of the credible intervals associated with some of the model parameters, including the \ac{MSP} mass and radius. 
We see this increase for both \texttt{ST-U} and \texttt{ST+PST} models. 
For the least computationally expensive model, \texttt{ST-U}, we investigated separately the effects of the different elements listed above. 
In Table \ref{tab:MR_STU_comparisons}, we show the $68\%$ credible intervals for mass and radius starting from what was reported in \citetalias{Riley2019} and ending with our reference results. 
In between we show the results of preliminary analyses that consequently implement the following changes\footnote{Each change is implemented in addition to the previous one(s).}, starting with the analysis of the old B19 data set and original response matrix, as used in \citetalias{Riley2019}, and finishing with the inference presented in the work for \texttt{ST-U} \NICER-only reference cases. 
\begin{description}[noitemsep]
    \item[test 1] new version of \XPSI, new modeling of uncertainties on instrument response, new priors;
    \item[test 2] the effect of different \MultiNest settings and in particular of increasing the number of live points;
    \item[test 3] the effect of changing the lowest considered channel from 25 to 30.
\end{description}
 Only the columns marked `reference' show results for the new data sets, analyzed with the corresponding updated \NICER response matrix. 

Adopting the new \XPSI framework, including the updated modeling of instrument uncertainties and priors, has the effect of shrinking the credible intervals when the same \MultiNest settings are used.  This can be explained by the fact that there are fewer parameters associated with the model. This reduces the space of possible solutions and hence the range of radii associated with them. 

Increasing the live points has the effect of increasing (asymmetrically) the credible interval. Although this was already pointed out for \joh in \citet{Riley2021, Salmi2022, Salmi2023}, 
for \joo this behaviour seems at odds with the findings of \citetalias{Riley2019}. There the authors performed three inference runs with a thousand live points. All of them generated well-overlapping contours, leading to the conclusion that a higher number of live points would not have influenced the results (unless other modes, with much smaller width but much higher likelihood were also present).  This seems not to be the case in the new framework for the \texttt{ST-U} model.  Similarly, \citet{Afle23} tested the dependence of the \texttt{ST+PST} model on the number of live points. When they increased the number of live points from 1000 to 4000, they found only a very small increase in the size of the credible regions (for instance the 1 $\sigma$ radius interval increased by 0.1 km).

Reducing the range of channels at low nominal energies also slightly widens the credible intervals. 
This highlights the important role played by events in the low energy tail of \NICER's sensitivity range, where the thermal emission from the hot spots is expected to to peak. Since with test 3, we are losing part of the information used previously, the widening is not surprising.  

Moving from the B19 to the updated B19v006 data set and the new instrument response slightly reduces the credible intervals once all of the changes mentioned above have been incorporated in the analysis framework. They are however overall still marginally wider than the original values published in \citetalias{Riley2019}. 

Comparing the \citetalias{Riley2019} results for the \texttt{ST+PST} model and the credible intervals of our new reference inference analysis, we find similar results, for likely similar reasons \citep[see also][]{Afle23}. 

In general, to obtain robust results in the new analysis framework and with the updated B19v006 data set, we require more computationally expensive \MultiNest settings (not all combinations have been tested, but e.g. adopting a larger number of live points seems necessary to obtain stable posteriors). 
 In particular our various \texttt{ST-U} and \texttt{ST+PST} analyses (including the ones presented in Section \ref{subsubsec:STU_ImpactSettings} and \ref{subsubsec:STPST-R19}) demonstrate that, in presence of a multi-modal structure in the posterior, adopting an insufficient number of live points has not only the  effect of underestimating the posterior widths, but it can also lead to significant biases. 
 
\begin{table*}[tbp]
\hspace*{-1.8cm}\begin{tabular}{c|c|c|c|c|c|c|c }
 & {\bf R19 \texttt{ST-U}} & {\bf test 1 \texttt{ST-U} }& {\bf test 2 \texttt{ST-U}}& {\bf test 3 \texttt{ST-U} }& {\bf reference \texttt{ST-U}}& {\bf R19 \texttt{ST+PST}}& {\bf reference \texttt{ST+PST}} \\
\hline
$R_{\mathrm{eq}}$\,[km] & 9.58 - 11.54 & 9.94 - 11.49 & 9.54 - 11.56& 9.68 - 11.91 & 9.64 - 11.68& 11.52 - 13.85 & 11.81 - 14.41 \\
\hline
$M\,[\mathrm{M_\odot}]$ & 1.02 - 1.20& 1.03 - 1.17& 1.023 - 1.199& 1.03 - 1.26 & 1.04 - 1.25 & 1.28 - 1.49 & 1.2 - 1.54\\
\end{tabular}
\caption{68\% credible intervals for mass and radius for different inference analyses adopting \texttt{ST-U} and \texttt{ST+PST} models, as expressed in the labels of the first row. In the labels {\bf R19} refers to the results reported in \citetalias{Riley2019}, {\bf reference} to the reference results presented in this paper and {\bf test 1-3}, to findings of preliminary analyses of the original data set used in  \citetalias{Riley2019}. {\bf Test 1-3} check for, in order: the effect of the new \XPSI framework and priors (for more details, see Section \ref{subsubsec:ParamPrior}); the effect of different \MultiNest settings (for more details, see Section \ref{subsec:multinest}); and finally, the effect of excluding channels [25,30).\\
Note that each change is implemented in addition to the previous one(s).}
\label{tab:MR_STU_comparisons}
\end{table*}

\subsection{Comparison with Findings from Simulations}
The simulation tests performed in \citetalias{Vinciguerra2023a} have highlighted expected and unexpected dependencies and behaviours of the results of our analyses. In this Section we comment on the most relevant ones, relating them to our findings based on the analysis of the B19v006 updated \NICER data set for \joo. 

\subsubsection{The Multi-modal Structure of our Posteriors}
In \citetalias{Vinciguerra2023a}, the authors revealed the presence of multi-modal structures in the posterior surface\footnote{We expect the presence of very similar modes and, in general, multi-modal structure also in the likelihood surface, given the mostly uninformative priors adopted for the \ac{PPM} analyses of \joo.}; this is common to both real data and simulations. 
 The presence of such structure is often clear only when the multi-mode/mode-separation setting of \MultiNest is activated or after inspecting tails of the posterior distributions. Due to the differences in posterior mass between the various modes, only for some cases the multi-modal structure is prominent enough to be perceivable by eye in the marginalized posterior distributions.
In both data and simulations, we clearly see very similar pulses and background generated by considerably different hot spot configurations, \ac{MSP} and interstellar properties. Multiple models and corresponding parameter vectors 
can reproduce \NICER data without
visibly impacting the quality of the residuals. 
This seems to differ from the case of \joh, where, with the tight mass, distance and inclination priors, a single and stable mode was identified, for both the \texttt{ST-U} and the \texttt{ST+PST} model, that was well-constrained in both the \ac{MSP} radius and in its hot spot geometry.

More and higher quality data, in addition to external constraints, could mitigate the challenges in the analysis of \joo's \NICER data set posed by 
this multi-modal structure of the posterior surface. 
For example, as we see when we introduce \xmm data, if we had a stringent and reliable background estimate for \NICER, that may lead solutions to a different portion of the parameter space and be less susceptible to multi-modality.

The various modes characterising the posterior surfaces often display different widths of the credible intervals. 
Moreover, as shown in \citetalias{Vinciguerra2023a}, posterior widths are quite sensitive to the specific noise realisation. 
It is therefore complicated to predict in advance the evolution of credible intervals once further data and constraints are included in the analysis. 
If however, the main identified mode is unchanged, 
we expect, with some scatter, that uncertainties in the 2D posterior distribution of mass and radius will decrease with exposure time or the square root of it, depending on how strong the correlation between the two is in that specific analysis \citep[see][]{Lo2013,Psaltis2014,Miller2015}. 

\subsubsection{Data versus Simulations}
In addition to the multi-modal structure, and just as in some of the analyses based on the synthetic data and reported in \citetalias{Vinciguerra2023a}, analysis of actual \NICER data also shows some variability in the results depending on the analysis settings. \\
However, comparing the results outlined in this paper with the findings from the simulations of \citetalias{Vinciguerra2023a}, we also notice some  differences. 
\begin{itemize}
    \item Here, as in \citetalias{Riley2019}, the inferred mass and radius change depending on the adopted hot spot model, while for the simulations, presented in \citetalias{Vinciguerra2023a}, 
     it did not matter whether the \texttt{ST-U} or \texttt{ST+PST} model was used: the resulting posterior distributions found with these two models for mass, radius and compactness were in any case very similar to each other. 
    \item Adopting the HR or LR \XPSI settings has a visible, although not significant, effect on the posterior distributions of mass and radius (as shown by comparing the green, HR, and the yellow, LR, lines in the left panel of Figure \ref{fig:STPST_res_sett}), difference that was not observed in the analyses of synthetic data presented in \citetalias{Vinciguerra2023a}. 
    \item Comparing the evidences obtained for 
    \NICER-only analyses when adopting different models, we see a significant preference for more complex hot spot shapes. 
    For the two cases tested in \citetalias{Vinciguerra2023a}, instead, this evidence difference reduced to a factor of a few in log-evidence, not enough to draw a decisive conclusion.  
\end{itemize}
The origin of this different behaviour for real data versus simulations could be due to a combination of several factors. 
The analyzed synthetic data may not cover a representative part of the parameter space; indeed in \citetalias{Vinciguerra2023a}, only one parameter vector per model was considered.  
If significantly more computational resources were available, it would be possible to produce and analyse multiple synthetic data sets, which would better sample the posterior inferred from real data (instead of limiting injected parameters to maximum likelihood samples). 
The different nature of the considered data may also generate some discrepancies in our findings. 
The simulations are based on synthetic data generated assuming the same physical processes and simplifications later adopted in the inference analyses. However, real data may contain physics that is not accounted for in our models: for example, hot spots are unlikely to have such simple shapes and will very likely display some temperature gradients. Normally we assume that our data are not sufficiently good for these pieces of missing physics to affect the outcome of the analysis; but this is something that should be tested in more detail.

{
    \begin{figure*}[t!]
    \centering
    \includegraphics[
    width=16cm]{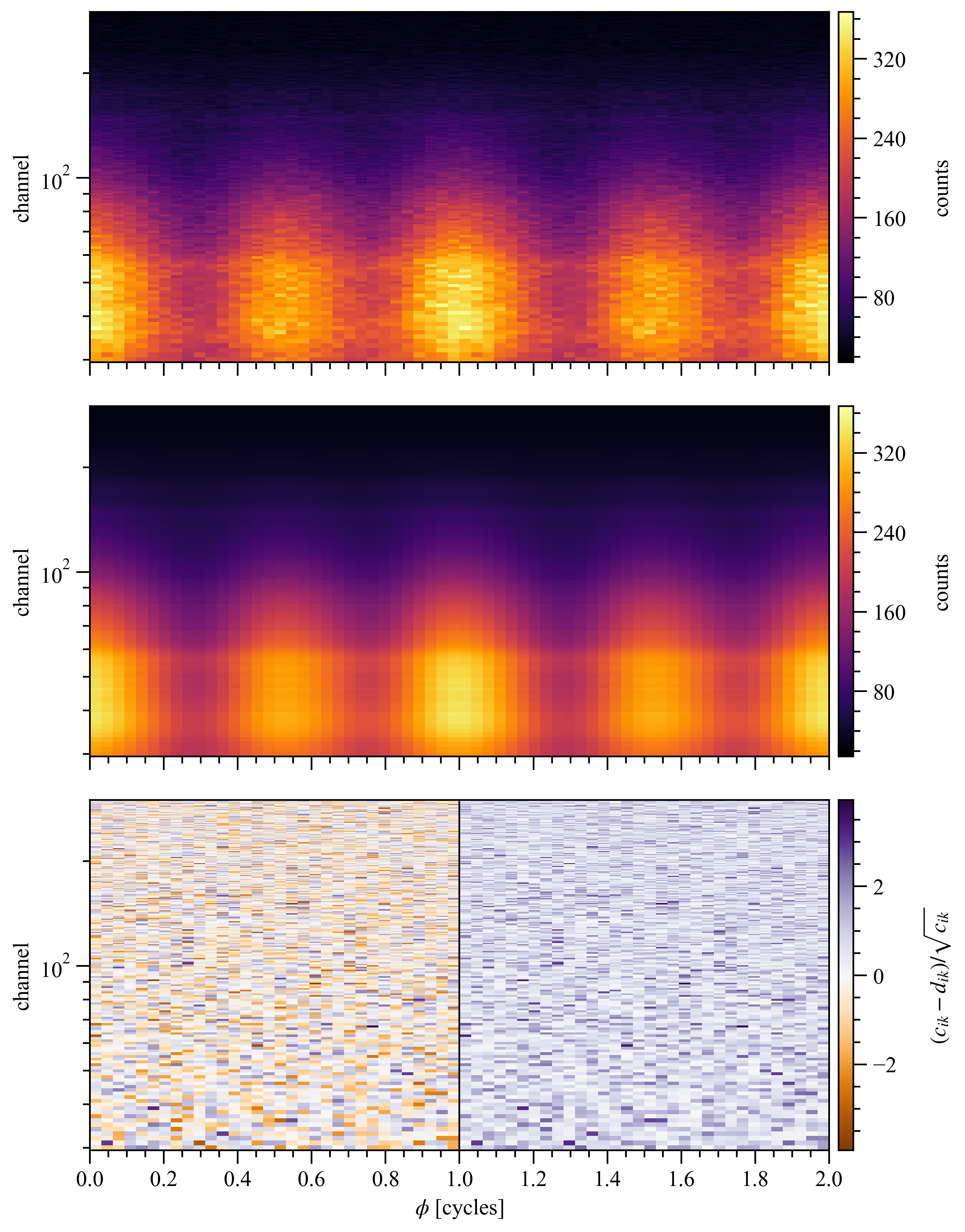}
    \caption{\small{
    The three panels represent from top to bottom: 
    data counts, posterior counts expected for the \texttt{PDT-U} model and residuals as a function of cycle phase (on the x-axis) and \NICER instrument channel (on the y-axis). 
    Residuals are defined as the difference in counts between data and expectations, weighted by the corresponding Poisson standard deviation (as expressed on the associated color axis):
    ${d_{ij}}$ are count data for the $i^{\mathrm{th}}$ phase bin and $j^{\mathrm{th}}$ channel, ${c_{ij}}$ the corresponding posterior-expected count numbers. 
    Expected counts are evaluated considering 200 posterior samples, inferred jointly analysing \NICER and \xmm data. 
    More details about this Figure can be found in the caption of Figure 13 and in Appendix A.2.1 of \citetalias{Riley2019}. 
    The complete figure set (8 images) is available in the online journal, for the \texttt{ST-U}, \texttt{ST+PST}, \texttt{ST+PDT}, \texttt{PDT-U} models for both \NICER-only analyses  (reference runs highlighted in bold in Table \ref{tab:runs})} and joint inferences for \NICER and \xmm data sets (analysis settings reported in Table \ref{tab:runs}, corresponding to ``yes'' entries in the \xmm column).
    } 
    \label{fig:data}
    \end{figure*}
}

\begin{table*}[tbp]
\hspace*{-2.5cm}\begin{tabular}{|l|l|l|l|l|l|l|r|}
\multicolumn{8}{c}{\text{\bf{\NICER-only}}} \\
\hline
{\bf Model}&{\bf Mass}\,$\mathbf{[\mathrm{M_\odot}]}$ & {\bf Radius}\,$\mathbf{[km]}$& $\boldsymbol{\log\left(\mathcal{E}\right)}$& $\boldsymbol{{\mathbf{\mathrm{max}}}\left(\log\left(\mathcal{L}\right)\right)}$ & $\boldsymbol{{\mathbf{\mathrm{max}}}\left(\log\left(\mathcal{L}_{\mathrm{N}}\right)\right)}$
& 
$\boldsymbol{{\mathbf{\mathrm{max}}}\left(\log\left(\mathcal{L}_{\mathrm{N\&XMM}}\right)\right)}$
& {\bf Core hours}\\
\hline
\texttt{ST-U} & $1.12^{+0.13}_{-0.08}$ & $10.53^{+1.15}_{-0.89}$ & -35788 & -35735& - & -42950 & 11462\\
\hline
\texttt{ST+PST}& $1.37^{+0.17}_{-0.17}$ &  $13.11^{+1.30}_{-1.30}$ & -35778& -35722 &  - & -42942 & 133970 \\
\hline
\texttt{ST+PDT}& $1.20^{+0.14}_{-0.11}$ & $11.16^{+0.90}_{-0.80}$ &-35779 &-35729 & - &  -42705 & 33069\\
\hline
\texttt{PDT-U}& $1.41^{+0.20}_{-0.19}$& $13.12^{+1.35}_{-1.21}$ & -35777& -35721 & - &  -42659 & 74333\\
\hline 
\multicolumn{8}{c}{\text{\bf{\NICER and \xmm}}} \\
\hline
{\bf Model}&{\bf Mass}\,$\mathbf{[\mathrm{M_\odot}]}$ & {\bf Radius}\,$\mathbf{[km]}$& $\boldsymbol{\log\left(\mathcal{E}\right)}$& 
$\boldsymbol{{\mathbf{\mathrm{max}}}\left(\log\left(\mathcal{L}\right)\right)}$
& 
$\boldsymbol{{\mathbf{\mathrm{max}}}\left(\log\left(\mathcal{L}_{\mathrm{N}}\right)\right)}$
& 
$\boldsymbol{{\mathbf{\mathrm{max}}}\left(\log\left(\mathcal{L}_{\mathrm{N\&XMM}}\right)\right)}$
& {\bf Core hours}\\
\hline
\texttt{ST-U}& $1.88^{+0.13}_{-0.19}$ &$15.12^{+0.64}_{-1.31}$ & -42714& -42661  & -35760 & -& 41219\\
\hline
\texttt{ST-U}$^{*}$& $1.35\pm 0.16$& $11.44\pm 1.14$ &-42718 & -42666 & -35769 & - & -\\
\hline
\texttt{ST+PST}& $1.93^{+0.10}_{-0.13}$ & $15.12^{+0.64}_{-1.31}$& -42711 & -42662  & -35761 & - & 84620\\
\hline
\texttt{ST+PDT}&  $1.40^{+0.13}_{-0.12}$ & $11.71^{+0.88}_{-0.83}$&-42684 &-42625 & -35730 & - & 3261\\
\hline
\texttt{PDT-U}& $1.70^{+0.18}_{-0.19}$ & $14.44^{+0.88}_{-1.05}$ & -42677 & -42617  & -35723 & - & 12421\\
\hline
\end{tabular}
\caption{
Summary of the results for the reference \NICER-only inferences and the joint \NICER and \xmm runs.
Uncertainties represent the 68\% credible intervals starting from the median of the marginalized 1D posterior distribution. 
Results above the double horizontal lines correspond to \NICER-only analyses; the ones reported below to joint \NICER and \xmm inferences.  For \NICER-only analyses, we report medians, credible intervals, evidences $\mathcal{E}$, maximum likelihood values $\log_{\mathrm{max}}\mathcal{L_{\mathrm{N}}}$ (in a joint \NICER and \xmm framework $\log_{\mathrm{max}}\mathcal{L_{\mathrm{N\&XMM}}}$) and core hours associated with the run with \MultiNest settings: SE 0.3, ET 0.1, LP $10^4$, MM on. 
The $\log_{\mathrm{max}}\mathcal{L_{\mathrm{N}}}$ marks the maximum likelihood values of the run identified by the line. 
The joint \NICER and \xmm inference runs are performed with \MultiNest settings: SE 0.8, ET 0.1, LP $10^3$, MM off, except for the \texttt{ST-U} case, the only model for which we could perform a SE 0.3, ET 0.1, LP $10^4$, MM on run. 
All the inference analyses, apart from those using the \texttt{ST-U} model, assume low resolution in terms of \XPSI settings (see Section 2.3.1 of \citetalias{Vinciguerra2023a}). For \texttt{ST-U} and \texttt{ST+PST} models, \citetalias{Riley2019} report respectively a mass of $1.09^{+0.11}_{-0.07}\, \mathrm{M_\odot}$ and $1.34^{+0.15}_{-0.16}\, \mathrm{M_\odot}$ and a radius of $10.44^{+1.10}_{-0.86}$\,km and radius of $12.71^{+1.14}_{-1.19}$\,km. \\
$^{*}$: Secondary mode of the \NICER and \xmm run with SE 0.3, ET 0.1, LP $10^4$, MM on (in this specific case, to describe mass and radius posterior, we use means and standard deviations and we report the value of the local evidence, as calculated by \MultiNest).
}
\label{tab:compare_models}
\end{table*}

\subsection{Model Comparison and the Effect of Including Constraints from \xmm}
\label{subsec:models} 
We report the main results for each model considered in this study in Table \ref{tab:compare_models}. 
The first four rows of the table refer to \NICER-only reference inferences (settings of reference runs are highlighted in bold in Table \ref{tab:MultiNestParams}); while the last five show results for joint \NICER and \xmm analyses (see specific settings in Table \ref{tab:MultiNestParams}, for entries with \xmm data included). 
From left to right, we report in each column:
(i) adopted model, 68\% credible interval for (ii) mass and (iii) radius, (iv) natural logarithm of  evidence, (v) natural logarithm of the maximum likelihood, (vi) only for joint \NICER and \xmm analyses -  natural logarithm of the maximum likelihood within the \NICER-only framework, (vii) only for \NICER-only inferences - natural logarithm of the maximum likelihood within the joint \NICER and \xmm framework (assuming $\alpha_{\mathrm{XTI}} = \alpha_{\mathrm{MOS1}} = \alpha_{\mathrm{MOS2}} = 1$, with $\beta$ defining the distance) and (viii) core hours necessary to compute the run.   
The second row in the joint \NICER and \xmm block refers to the secondary \texttt{ST-U} mode, identified in the \texttt{ST-U} inference run, whose main results are summarized in the row just above. This mode is particularly interesting since a similar configuration is also found adopting the \texttt{ST+PDT} model and it resembles the main findings for \joh. In this case only, in the fourth column, we report the local evidence; for all other rows, we refer to the global evidence. 

We highlight the implications of our findings, summarized in Table \ref{tab:compare_models}, in the following subsections. 

\subsubsection{NICER-only Analyses} 
\label{subsubsec:discussion_NICERonly}
The first four rows of Table \ref{tab:compare_models} summarize the results of the \NICER-only inference runs. 
    As in \citetalias{Riley2019}, the evidence clearly disfavours the \texttt{ST-U} model, but does not display a clear preference for any of the more complex models.
    However, while the analysis settings for all \NICER-only runs whose results are displayed in the table
    are the same, this does not necessarily imply the same 
    accuracy in the evidence evaluation, since the dimension of the parameter space increases with the complexity of the model. Hence, how well each model explored the parameter space would ideally need to be checked in more detail to corroborate our findings. 
    Note for example that the 
    maximum likelihood value found for the \texttt{ST+PST} model is higher than the one identified in our reference \texttt{ST+PDT} inference run. Since the \texttt{ST+PST} model is approximately contained in the more complex \texttt{ST+PDT} model, this likely implies that our reference \texttt{ST+PDT} run has not adequately explored the parameter space\footnote{Another possibility could be that the prior mass (note that this is not the same as the posterior of the pulsar mass parameter, or the mass posterior) of this mode is too small for \MultiNest to consider it relevant for the evidence estimation. This latter hypothesis is however unconvincing, since a detailed look into the posterior samples showed no solution similar to the one found in \texttt{ST+PST} and hence the total contribution of this mode has at least not been accurately estimated by \MultiNest.}. This has  implications for the estimated evidence, whose value is therefore likely underestimated.

    Exclusively in terms of mass and radius posteriors, 
    two main classes of solutions are present:  $M\sim 1.1-1.2\,\mathrm{M_\odot}$ and $R_{\mathrm{eq}}\sim 10.5-11.1\,$km; or $M\sim 1.4\,\mathrm{M_\odot}$ and $R_{\mathrm{eq}}\sim 13\,$km \footnote{These two groups of mass and radius values, do not however reflect the hot spot configurations associated with them, which show a greater variety.}.
     The former seems to be slightly more asymmetric and narrower than the latter (given the same adopted settings, which however again do not guarantee the same quality of the uncovered posterior).

   The newly introduced \texttt{ST+PDT} and \texttt{PDT-U} models, which were not studied in \citetalias{Riley2019}, show significant posterior mass for configurations with at least one of the hot spots on the same hemisphere as the observer.  These geometric configurations were not found in \citetalias{Riley2019}.  

\subsubsection{Joint \NICER and \xmm Analyses}
\label{subsubsec:discussion_NICERXMM}
We focus now on the entries of Table \ref{tab:compare_models} corresponding to joint \NICER and \xmm analyses.

\begin{description}
\item [\bf{Evidence}] 
unlike for the \NICER-only analyses, for the joint inferences, the evidence indicates a significant preference for the \texttt{PDT-U} model. 
Both \texttt{ST-U} and \texttt{ST+PST} models are strongly disfavoured, with similar evidence values differing from the best one by factors of $\sim 40$ in $\log$ units. 
The difference between \texttt{PDT-U} and \texttt{ST+PDT} is milder (about $7-8$ in $\log$ units), but still decisive. 
If we assume that including \xmm data introduces unbiased constraints, the substantial shift in solution(s) driven by the additional \xmm data, compared to those reported in \citetalias{Riley2019}, 
shows that the \joo ~\NICER results need to be interpreted carefully. 
This implies that, for this source, whose data are clearly characterised by a prominent multi-modal structure, further constraints could once again significantly modify the results of our inference. 

    \item [\bf{Radii inferred with \texttt{ST-U} and \texttt{ST+PST} models}]    for the \NICER and \xmm joint analyses, the new main solutions identified with the \texttt{ST-U} and \texttt{ST+PST} models are very similar to each other, in mass and radius as well as in hot spot configurations. 
    Some of the similarities, especially in terms of mass and radius, could be caused by the radius posteriors' tendency of reaching the upper limit of the prior and consequently constraining the mass through inferences on the compactness. 
    We also highlight that such high radius values are
    difficult to reconcile with constraints posed by other observations, such as the inferred tidal deformability from gravitational waves \citep[see e.g.][]{Abbott2018,Abbott2019} and from the analysis of \NICER (and \xmm) \joh data \citep{Miller2021, Riley2021,Salmi2022}. 

\item [{\bf Hot spot complexity}]
introducing the masking component that distinguishes the \texttt{ST+PST} model from the simplest \texttt{ST-U} one does not improve our ability to represent both \NICER and \xmm data. 
 Looking at the corresponding geometrical parameter posterior distributions (see Figure set \ref{fig:params} in the online journal) there is a clear degeneracy due to the small size of both hot spots (degeneracy of type I, as described in \citetalias{Riley2019}). This generates ambiguity in the association of the \texttt{ST} and \texttt{PST} labels with the hot spots (i.e. either hot spot could in principle be represented with or without the omitting component). 
On the other hand, when we introduce an emitting component via the \texttt{ST+PDT} model, its association with the hot spot which we see first in the data is strongly favoured, and the associated evidence and maximum likelihood values improve drastically. 
Introducing a second emitting component also for the other hot spots, as present in the \texttt{PDT-U} model, further improves the quality of our results. 
 \item [{\bf Similarities in the tails of the posteriors}] further examinations of the posterior distributions derived with the \texttt{ST+PST} model reveal that the configurations populating the tails at low radii are likely forming a secondary mode, as was the case for the joint \NICER and \xmm \texttt{ST-U} analysis. 
These secondary modes display similar geometries (being viewed almost from the equator and with almost antipodal hot spots), masses, and radii as the main \texttt{ST+PDT} mode. 
As we will discuss further in Section \ref{subsubsec:discussion_gamma}, this configuration is of particular interest because it could potentially better explain the gamma-ray emission associated with \joo; 
moreover, it would be similar to the geometry found for \joh (which we may expect if the underlying magnetic field and its history are similar).  
\end{description}

\subsubsection{Transitioning from \NICER-only to Joint \NICER and \xmm Analyses}
\label{subsubsec:discussion_overall}
In this Section we briefly outline the main effects that we observe in our findings, transitioning from analysing only the \NICER data to jointly considering both \NICER and \xmm data sets. 
We however remind the reader that, as more extensively discussed in Section \ref{subsec:caveats}, inferences of \NICER-only data were performed with better (and more computationally expensive) \MultiNest settings, compared to the joint instrument analyses (the joint analyses generally required a larger number of core hours, given the same settings, to reach the termination condition set by the evidence tolerance). 

We start by noting that possible differences in the results were already anticipated in \citetalias{Riley2019} (see the discussion in Section 4.1.2 of that paper). 
Introducing \xmm data has the general effect of decreasing the inferred background, leading to higher compactness values \citep[following the same logic presented in][]{Riley2021,Salmi2022}. This is in general true for all models, which, although in different ways, always reach higher compactness by shifting the posteriors of both masses and radii toward higher values. 
However, the findings obtained adopting the \texttt{ST-U} and \texttt{ST+PST} models are the ones mostly affected by the change. 
Indeed the \texttt{ST-U} and \texttt{ST+PST} \NICER-only analyses identify, as main modes, solutions that are only supported by higher background levels, compared to the broader background support for the \texttt{ST+PDT} and \texttt{PDT-U} main modes. 

For the \texttt{ST+PDT} and \texttt{PDT-U} models, introducing \xmm data into the analyses has the effect of reducing the background support for the main modes identified with \NICER-only inferences, whilst still maintaining them as the most probable configurations
(now characterised by slightly narrower posteriors, looking e.g. at radius and mass, as a consequence of the restricted background support of the same mode)\footnote{Again, however, the main mode, identified by the \texttt{ST+PDT} model in \NICER-only analyses, is very unlikely to be the one containing the highest likelihood values (see reasoning in Section \ref{subsubsec:discussion_NICERonly}).}.  

Instead, solutions found with \texttt{ST-U} and \texttt{ST+PST} in the \NICER-only inferences are downplayed by the introduction of the constraints posed by \xmm data, which jointly with \NICER data can now be better represented by a completely different mode. 
As mentioned in Section \ref{subsubsec:discussion_NICERXMM}, both \texttt{ST-U} and \texttt{ST+PST} models favour solutions with high radius values ($R_{\mathrm{eq}}\sim15-16$\,km), and hot spot configurations resembling panel C of Figure \ref{fig:config}, for the main bulk of the posteriors. 

Introducing \xmm data and constraints does not significantly affect the ratio between $\alpha_{\mathrm{XTI}}$ and the distance squared, or the $\beta$ parameter in \NICER-only analyses (representing the only distance and $\alpha_{\mathrm{XTI}}$ combination to which the analysis is sensitive), which remains between $\sim8.4-8.9$.  

Transitioning to joint analyses has different effects on the inferred column density, depending on the assumed model. This is most likely an indirect consequence of changes (for \texttt{ST-U} and \texttt{ST+PST}), or the absence of changes (for \texttt{ST+PDT} and \texttt{PDT-U}), in the identified main mode. 
For both \texttt{ST-U} and \texttt{ST+PST} the inferred $N_{\mathrm{H}}$ posterior shifts from peaking at specific values ($\sim 1.48\times 10^{20}$cm$^{-2}$ and $0.9\times 10^{20}$cm$^{-2}$) to be completely consistent with zero (i.e. clearly limited at $\sim 0.1\times 10^{20}$cm$^{-2}$ by the lower limit of our prior).  
For the \texttt{ST+PDT} analyses, introducing \xmm data does not significantly affect the $N_{\mathrm{H}}$ posterior, which clusters around $0.7\times 10^{20}$cm$^{-2}$. 
Similarly for \texttt{PDT-U}, the joint analysis points towards an inferred $N_{\mathrm{H}}$ posterior only marginally different from what was found in \NICER-only inferences, with median values changing from $1.58\times 10^{20}$cm$^{-2}$ to $1.48\times 10^{20}$cm$^{-2}$. 
Given the different estimates of $N_\mathrm{H}$ that the various models provide, it is clear that 
strong constraints would considerably help our analyses and improve our interpretation of their results. 
In future analyses, we plan to include a more realistic prior on the $N_\mathrm{H}$, based on literature. We  hope to find up-to-date estimates in particular of its lower limit, as this would greatly benefit our analyses. 

Finally, introducing \xmm does not seem to significantly affect the quality of the \NICER residuals (which are still good, see Figure set \ref{fig:data} in the online journal), while the \xmm data are also well represented (for an example, see panels in Figure \ref{fig:BKG_MOS1}). 
From Table \ref{tab:compare_models}, we can see that the maximum likelihood for \NICER-only drops considerably if we use the maximum likelihood solution identified in the joint analysis for the \texttt{ST-U} and \texttt{ST+PST} models (with a drop of about 30 in $\log$ likelihood). 
However, this is not the case for \texttt{ST+PDT} and \texttt{PDT-U}. 
The former shows a similar maximum likelihood value to that found in the corresponding \texttt{ST+PDT} \NICER-only analysis; while for the latter the best fitting solution has a \NICER likelihood comparable to the overall best solution from the \NICER-only inferences (and therefore very close to the one identified with the \texttt{ST+PDT} \NICER-only analysis)\footnote{Here we compare maximum likelihood values to have a rough estimate on what is the worsening in matching the data, required to include the \xmm constraints.}. 

\subsubsection{New versus Old Favoured Configurations}
As mentioned in Section {\ref{subsubsec:discussion_NICERXMM}}, overall we find that the \NICER and \xmm data sets can be jointly well described by the solutions identified with both \texttt{ST+PDT} and \texttt{PDT-U} models. 
Interestingly, compared to what was found for the models and data analyzed in \citetalias{Riley2019}, in neither case there is a need for elongated hot spots or for both the hot spots being located in the same hemisphere (see panels E and F of Figure \ref{fig:config}). 
For both models, the {\texttt{PDT}} hot spot(s), i.e. the ones described by two emitting elements, are characterised by a larger cooler component and a much hotter and smaller one. 
This hints at the presence
of temperature gradients, to which our analysis is apparently sensitive. We believe this should motivate investigations aimed at characterising the sensitivity and robustness of our findings to more realistic physical properties of the \ac{MSP} surface hot spots. 
We also note that for both models, the overall dimensions of the two hot spots are comparable and that they are almost antipodal (at least in phase). 
These characteristics resemble the configuration found by analysing \NICER-only observations and jointly the \NICER and \xmm data for \joh. 
In particular the solutions identified with \texttt{ST+PDT} show even more similarities with the findings for \joh, including the high inclination ($\cos(i) \sim 0.1$) and the relatively low radius value of $R_{\mathrm{eq}}\sim 12$\,km.  As mentioned previously, very similar solutions are also found in the tail of the joint \NICER and \xmm posteriors derived with the \texttt{ST-U} and \texttt{ST+PST} models, as secondary modes.

\subsubsection{New Favoured Configurations in Relation to Gamma-Ray constraints}
\label{subsubsec:discussion_gamma}
The hot spot configurations found with \texttt{ST+PDT} and as a secondary mode of the \texttt{ST-U} (and \texttt{ST+PST}) joint \NICER and \xmm inferences are, interestingly, more consistent with the gamma-ray emission associated with \joo.  \citet{Kalapotharakos2021} considered multipolar field geometries consistent with the mass, radius, and viewing geometry (inclination $\sim50$ degrees) results of \citetalias{Riley2019} and \citet{Miller2019}, under the assumption the hot spots correspond to open field line regions. Open field lines were identified by integration from the surface to the light cylinder in either static vacuum or force-free models, with corresponding identified surface regions assumed to be uniform in temperature. Only bolometric light curves were considered in \citet{Kalapotharakos2021}. The offset dipoles and quadrupoles resulted in a variety of surface patterns qualitatively different in shape but consistent in position with \citetalias{Riley2019} and \citet{Miller2019}’s highest likelihood shapes. 

However, \citet{Kalapotharakos2021} also found degeneracies, i.e. a multimodal likelihood surface, when only the X-rays were considered. A key aspect of the force-free models of \citet{Kalapotharakos2021} was that they allowed for the computation of gamma-ray emitting regions, beyond the light cylinder in the current sheet. This high-altitude field region's morphology is most sensitive to the global dipolar field geometry and orientation relative to the spin axis. A key novel aspect in \citet{Kalapotharakos2021} was the computation of synchronous (in absolute phase, relative to the radio) X-ray and gamma-ray pulse profiles with multipolar fields — this involved correcting for light travel times and different emitting regions in curved space-time. Of the force-free models considered in \citet{Kalapotharakos2021}, only one (model $\mathrm{RF4\_{11}}$) seems to describe both the X-rays and gamma-rays well in phase alignment. This model’s X-ray hot spots are circular and are reminiscent of models and modes D and E of Figure \ref{fig:config} in this work. The study by \citet{Kalapotharakos2021} implies that a self-consistent treatment of both X-rays and gamma-rays will reduce parameter degeneracies and identify the correct hot spot modes and viewing geometry for \joo. Such an investigation is planned. Note that seeing both radio and gamma-rays from a pulsar generally limits the viewing geometry (the radio is from magnetic polar regions, while the gamma-rays are from equatorial current sheets). These visibility regions cut across almost orthogonally in the space of dipole inclination angle and spin axis viewing direction \citep[see for example][]{Corongiu2021}.

{
    \begin{figure*}[t!]
    \centering
    \includegraphics[width=8.9cm]{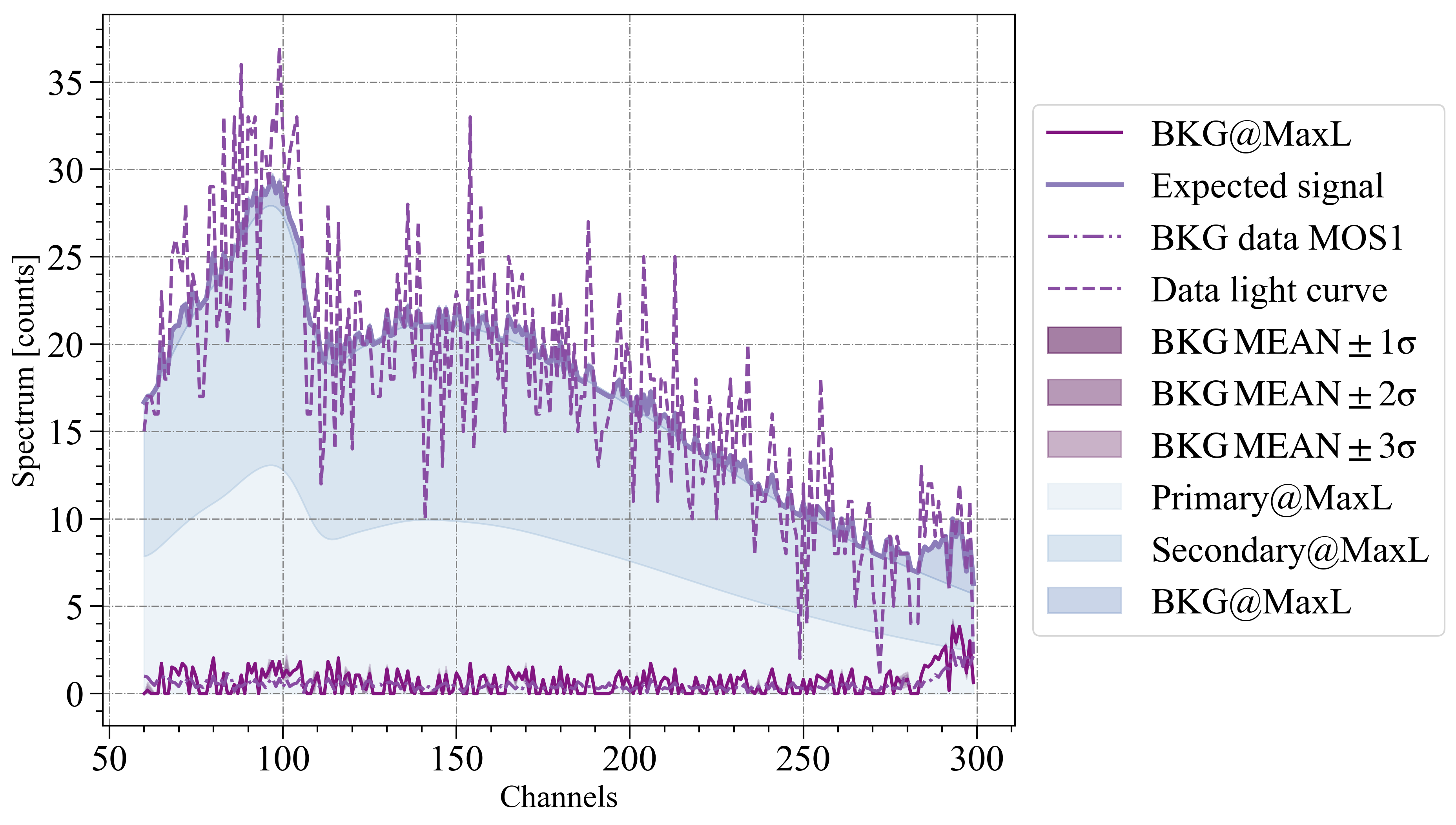}
        \includegraphics[width=8.9cm]{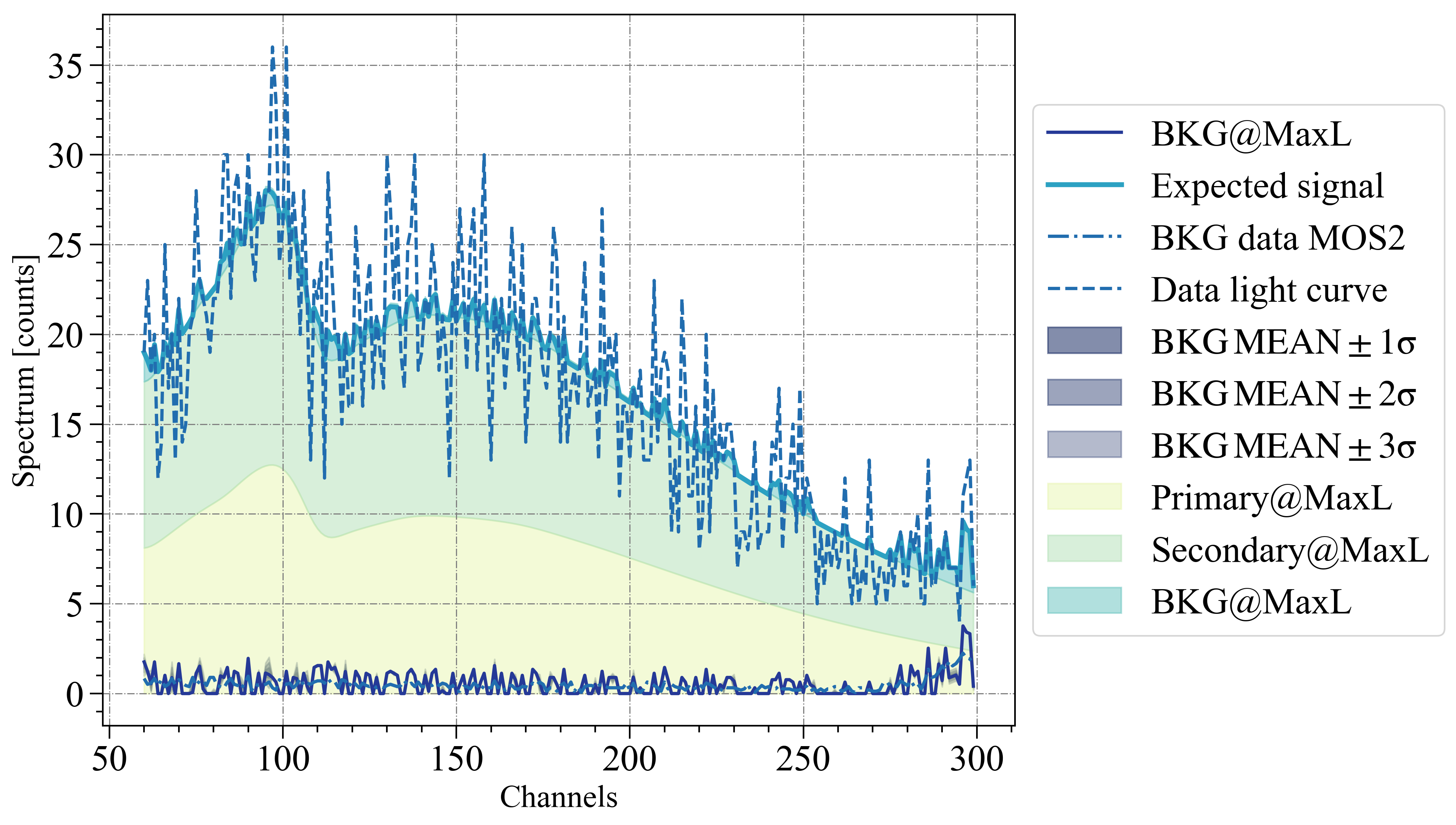}
    \caption{\small{
    Example showing how our \texttt{PDT-U} maximum likelihood (MaxL, in the legend) solution, obtained in the joint \NICER and \xmm analysis,  represents \xmm \joo and background (BKG) data, for MOS1 (left panel) and MOS2 (right panel).
    The plot shows similar quantities as Figure \ref{fig:BKG}, but for the \xmm MOS1 and MOS2 total and background data sets. 
    The dash-dotted line (magenta, for MOS1 and light blue, for MOS2) shows the background data used in our analysis; the inferred background is plotted with a solid line (in magenta for MOS1 and dark blue for MOS2). Given its low value, the uncertainties (shown with shaded areas, purple for MOS1 and dark blue for MOS2) are almost imperceptible. 
    The dashed more scattered line shows the total on-source data, for MOS1 on the left and MOS2 on the right; the  solid line (purple for MOS1 and blue for MOS2) is the inferred expected signal. 
    We display the contribution of the primary, the secondary, and the background associated with the maximum likelihood sample, adding them (in this order) on top of each other with areas of different intensities of blue, for MOS1, green for MOS2.
    } 
    }
    \label{fig:BKG_MOS1}
    \end{figure*}
}

\subsection{Caveats}
\label{subsec:caveats}
In this Section we list the major caveats and assumptions that affect the analyses presented here. 
\begin{description}
    \item[{\bf Atmosphere state and composition}] for the analyses in this paper we have assumed a fully ionized hydrogen atmosphere. While this assumption is quite standard, and well-motivated, it can have a substantial impact on the parameter posteriors that we recover, as discussed and described in detail in \citet{Salmi2023}.
    \item[{\bf Simplification of the hot spot properties}] 
    here, as in all \NICER studies so far, we adopt models that greatly simplify the physics characterising the hot spots. The underlying assumption is that our data and relative uncertainties are not sensitive enough to such details for us to need to account for them. 
    Degeneracies, in the analyses presented both here and in  \citetalias{Vinciguerra2023a}, show that we are not sensitive e.g. to the direction of arcs, if thin and elongated, or even to the presence of a structure more complex than a circle, if the hot spot is small enough (see e.g. label ambiguity in the \texttt{ST+PST} joint \NICER and \xmm analysis). 
    However, the jump in evidence and maximum likelihood values obtained when introducing a second emitting component per hot spot suggests that our analyses are in fact sensitive to temperatures and, likely, temperature gradients (considering indeed the usually very small size of the hotter component). 
    \item[{\bf Priors}] 
    priors are particularly important in our inference analyses, as  they drive the \MultiNest exploration of the parameter space and sometimes they clearly impact the range of our credible intervals. That is for example the case for the radius posteriors inferred in the \texttt{ST-U} and \texttt{ST+PST} joint \NICER and \xmm analyses, where our hard upper limit clearly modifies the natural 1D posterior distribution of that mode, truncating it at 16\,km and thus skewing the higher radius end of the credible interval. 
    Priors also play a crucial role in the evidence and yet their definition always, at least partially, relies on arbitrary choices: firstly through the chosen parametrization to describe the problem and secondly through their practical implementation, which often includes approximations, simplifications and arbitrary cut-off values. Thankfully, when the posterior is data-dominated, the impact of these choices is heavily suppressed. However, any time that we use the evidence, we should keep in mind that priors are crucial to evaluate it. As noted, our reference \texttt{ST+PDT} in \NICER-only analysis missed exploring an important part of the parameter space, since the solution for \texttt{ST+PST} is also in the \texttt{ST+PDT} parameter space and has a better likelihood. However, it is also possible that \MultiNest did not pick such a solution because its supporting prior volume is too small (if for example it required temperatures very close to the prior lower bounds) when put in the higher dimension parameter space describing the \texttt{ST+PDT} model. It should be stressed that 
    \MultiNest explores the parameter space with the aim of computing the evidence, through definitions of iso-likelihood contours \citep{Feroz2008,Feroz2009,Feroz2019}. 
    \item [{\bf Use and interpretation of \xmm data}] introducing \xmm data could lead to biases if the underlying assumptions are not correct, e.g. if the assumption on cross calibrations are incorrect, 
     or other astrophysical contamination is present at the pulsar sky location without being accounted for in our models.
     
    However, while we should keep this in mind and test for alternatives when possible, we think it is a crucial procedure
    to cross-check the compatibility of our findings with other observations, given the large amount of data increasingly available. 
    Hence, independently of the outcome, we believe including \xmm data and constraints promotes correct scientific practice. 
    Cross checks are especially important for \NICER sources like \joo, which lack independent information to coherently incorporate in priors (unlike for \joh, 
    and other \NICER sources like \joA, 
    where radio constraints can considerably inform our analyses and reduce the prior parameter space associated with masses and inclination).
    \item[{\bf Adequacy of the analysis settings}] 
    The data quality for \joo is very high, which increases the computational cost of this type of analysis\footnote{For the same settings, the analysis of \joh ~\NICER data, which have a much lower quality than for \joo, required considerably fewer core hours. This is one of the reasons why that source has been analyzed with better \MultiNest settings; a much larger amount of live points and a lower sampling efficiency.}.  Given the computational resources available to us, we could not fully test how well the parameter space was explored for the most complex models; analyses adopting \texttt{ST-U} and focusing on \NICER-only data were the only exception.  With the various tested settings, it was possible to prove that in this case our reference run had reached a converged state (although a significantly higher number of live points could potentially still reveal unexplored and relevant portions of the parameter space). The exploratory runs, which adopt an increasingly higher number of live points, also proved that merging the results of multiple inference runs with lower live points is not always equivalent to a single run with higher live points (e.g. merging all our \texttt{ST-U} runs with LP $10^3$ would have converged to a radius about $\sim1\,$km higher than what we find with a more adequate amount of live points $(6-10)\times 10^3$).  
    Indeed, contrary to expectations, we find, as also shown to a lesser extent for \joh in \citet{Riley2021,Salmi2022}, that a lower number of live points does not only underestimate the width of a mode, but also slightly biases it. In our \texttt{ST-U} \NICER-only cases, for example, we see a systematically higher median values when only $10^3$ live points are used. However, further checks are still needed; e.g. it is yet unclear if decreasing the evidence tolerance to 0.001 would have an effect when a high number of live points is employed. For the more complex models, we cannot guarantee an exhaustive exploration of the parameter space given its increased dimension and 
    the hence increased computational cost\footnote{This implies that we cannot guarantee that significant posterior shifts and changes would not appear in case of repeated inference runs with the same analysis settings.}. 
    This is particularly true for the joint \NICER and \xmm inference, whose requirements in terms of analysis settings had to be lowered to render them computationally tractable.  
As suggested in \citetalias{Vinciguerra2023a}, future analyses of upgraded data sets, targeting a new headline inference of mass and radius for \joo, will need adequate studies to address whether settings are appropriate, given the data and the model framework. 
More tailored techniques may also be employed to assess robustness in the obtained posterior and the adequacy of the adopted models. 
The robustness of the findings for the \texttt{ST+PST}, \texttt{ST+PDT} and \texttt{PDT-U} models, both for \NICER-only inferences and joint \NICER and \xmm analyses, needs more extensive testing and this will be considered in our computational budget requests for future analysis. 

    \item[{\bf Correct interpretation of evidences}] 
    in the presence of a highly multi-modal structure of the posterior surfaces, it is important to reflect on what should be inferred from evidence values. 
    In particular, 
    if one model is preferred by the evidence, it does not necessarily imply that the radius posterior associated with its main mode is also favored. 
    We also point out that, in this context, preferences in terms of posteriors seem to be naturally driven by their correspondent local evidence. 
    However this is also an arbitrary choice. 
    Moreover the evidence depends on the priors which, as we mention above, are defined with some arbitrary components. Therefore, in certain contexts, it may be preferable to focus instead on the best fitting solution (i.e. the solution given the highest likelihood or posterior value), independently from its width, instead of the more ``robust" solution, found with evidence and hence backed up by a larger volume in the prior space.  
    
\end{description}

\section{Conclusions}
\label{sec:conclusion}
In this paper we analyzed the reprocessed \NICER data set for \joo, B19v006. 
We adopted four different models to describe increasing complexity in the hot spot shapes: \texttt{ST-U}, \texttt{ST+PST}, \texttt{ST+PDT} and \texttt{PDT-U}. 
The former two were both found in \citetalias{Riley2019} to well represent the data (i.e. they had no suspicious structure in the residuals); however the more complex model \texttt{ST+PST} was favoured by the evidence. 
According to this model, \joo is a \ac{MSP} characterised by a mass of about $\sim 1.3\,\mathrm{M_\odot}$ and a radius of $\sim 13$\,km. Both hot spots were located in the same hemisphere, opposite to the observer, and one of them had an elongated shape. 
\texttt{ST+PDT} and \texttt{PDT-U} are two additional models tested for the first time for \joo. They allow for an additional emitting component to describe one, for \texttt{ST+PDT}, or both of the hot spots, for \texttt{PDT-U}.

In this work we upgraded the data set release that was described in 2019 using an updated instrument response, and also coherently included \xmm data in our analyses. 
Compared to \citetalias{Riley2019}, we here also adopted more extensive analysis settings in our main inferences. However, we were able to check the robustness of the obtained posterior distributions only for the simplest \texttt{ST-U} model. 
The computational resources necessary to test the adequacy of the analysis setting for inferences adopting more complex models were instead prohibitive. 
With this caveat in mind (see also Section \ref{subsec:caveats}), below we summarize the main takeaway points of our inferences. 
\begin{itemize}
    \item Analysing the updated \NICER data set, with new settings and an upgraded pipeline, we produce results consistent with what was found and reported in \citetalias{Riley2019}.
    \item Based on \texttt{ST-U} tests, compared to the analyses of \citetalias{Riley2019}, the reprocessed data and new framework seem to require a larger number of \MultiNest live points to ensure the robustness of the inferred posterior distributions.
    \item Introducing \xmm, and hence indirectly incorporating background constraints,  considerably affects the \joo findings.
    \item Joint analyses can require a considerable amount of computing resources while reducing \MultiNest and \XPSI settings to make the analysis computationally tractable can have significant effects on the final findings. 
    To produce robust results, analyses of future data sets will require the allocation of more computational resources and/or a considerable speed-up of the parameter estimation procedure. 
    \item In joint \NICER and \xmm inferences, \texttt{ST-U} and \texttt{ST+PST} give rise to very similar solutions (but very different from the corresponding ones identified with \NICER-only data), which lead to very large masses and very high radii (which would be in strong tension with the results from GW170817), the latter hitting the edge of our hard-cut prior at 16\,km.
    \item Both \texttt{ST+PDT} and \texttt{PDT-U} models, with the latter being significantly preferred by the evidence, are able to reproduce both \NICER and \xmm data, without invoking elongated shapes for hot spots or requiring them both to be located in the same hemisphere. The hot spot configurations found with these two models are likely in better agreement with the \joo gamma-ray emission and resemble the findings of \joh, especially for the solutions associated with the \texttt{ST+PDT} model. Radii and masses estimated during the preliminary joint \NICER and \xmm runs, presented in this work, for \texttt{ST+PDT} and \texttt{PDT-U} are respectively: $R_{\mathrm{eq}}\sim 11.5$\,km, $M\sim 1.4\,\mathrm{M_\odot}$ and $R_{\mathrm{eq}}\sim 14.5$\,km, $M\sim 1.7\,\mathrm{M_\odot}$. 
    However, more targeted and detailed studies are needed to assess the robustness of these findings.
    \item The configurations identified by the \texttt{ST+PDT} and \texttt{PDT-U} models, and how well they seem to describe the data, suggest the presence of gradients of temperature in the hot spot. It is therefore crucial to check what would be the impact of such gradients on our parameter recovery, through more realistic simulations. 
    \item Our findings also suggest that our analysis is only weakly sensitive to small details describing the hot spots 
    (e.g. for the secondary mode found for the \texttt{ST+PST} \NICER-only analyses, the omitting component can assume different sizes and location, as long as the area of the effectively emitting regions maintains a similar size).
    \item Different models and parameter vectors are currently in good agreement, given the quality of the residuals, with the B19v006 data and constraints. 
    If this situation does not change in the future, it could be useful to provide a more informative (constrained) prior for \joo's radius, based on other robust observations, such as estimates of the tidal deformability from gravitational waves and/or constraints from other \NICER sources. One could also develop more adequate procedures to deal with multi-modal solutions for follow-on equation of state studies. 
    \item It appears that more factors complicate the analysis of \NICER \joo data compared to \joh. One reason could be \joh's high inclination and general intrinsic hot spot configuration. It could also be due to the quality of the data, since for \joh (which is faint in X-ray) we know that most \NICER photons actually come from the background. The tight constraints on mass and inclination coming from radio observations, thanks to \joh's parameters and geometry, are also extremely helpful for the analysis of that source.
    \item Constraints from other studies and observables would be very useful, for example from \joo's gamma-ray pulse profile or \NICER background estimates. 
\end{itemize}
\subsection{Summary}
In summary, we analyzed the \joo data set, described in \citet{Bogdanov19a}, with the most updated instrument response and found, with an updated analysis and modeling framework, results compatible with the findings reported in \citetalias{Riley2019}. 
Our in-depth analysis, however, reveals a complicated structure of the likelihood surface that is difficult to explore extensively, given our current computational resources. 
We introduce constraints posed by the \xmm X-ray observations of \joo. 
This suggests that there may be no need for both hot spots to be located on the same hemisphere, nor for elongated shapes, as long as multiple temperatures can be assigned to a single hot spot. 

Given the inference analyses conducted for this work (and noting the caveats outlined in Section \ref{subsec:caveats}), there is a significant preference, in terms of evidence values, for the most complex model tested (\texttt{PDT-U}), which can seemingly well represent both \NICER-only and joint \NICER and \xmm data. The radius and mass associated with this solution are: $14.44^{+0.88}_{-1.05}$\,km and $1.70^{+0.18}_{-0.19}\, \mathrm{M_\odot}$. However, we find that \texttt{ST+PDT} also well reproduces both \NICER-only and joint \NICER and \xmm data. It does so with a hot spot configuration that resembles \joh, and radius estimation is more in line with the value currently inferred for that source. The \texttt{ST+PDT} model yields a radius and mass of $11.71^{+0.88}_{-0.83}$\,km and $1.40^{+0.13}_{-0.12}\, \mathrm{M_\odot}$ for \joo.

This in-depth re-analysis of the 2017-2018 \NICER data set indicates that \joo is a complex source to model.  Due to the multi-modal nature of the solution space, it is a source for which additional constraints - on the background, or on the system geometry - will be important in determining a robust inference of mass and radius.  This study provides a more comprehensive baseline to support in-progress analysis of new, larger \NICER data sets for this source, including benchmarking of the computational resources that need to be allocated to ensure robust results.

\begin{acknowledgments}
The authors thank Evert Rol and Martin Heemskerk
for technical assistance. 
This work was supported in part by NASA through the \NICER mission and the Astrophysics Explorers Program. S.V., T.S., A.L.W., D.C., Y.K. and T.E.R. acknowledge support from ERC Consolidator Grant No.~865768 AEONS (PI: Watts).  This work was sponsored by NWO Domain Science for the use of the national computer facilities.  NICER work at NRL is supported by NASA. S.B. acknowledges support through NASA grant 80NSSC22K0728. SG acknowledges the support of the CNES. W.C.G.H. acknowledges support through grant 80NSSC23K0078 from NASA. S.M.M. acknowledges support from NSERC Discovery Grant RGPIN-2019-06077. A portion of the results presented was based on observations obtained with XMM-Newton, an ESA science mission with instruments and contributions directly funded by ESA Member States and NASA. This research has made use of data and software provided by the High Energy Astrophysics Science Archive Research Center (HEASARC), which is a service of the Astrophysics Science Division at NASA/GSFC and the High Energy Astrophysics Division of the Smithsonian Astrophysical Observatory. We acknowledge extensive use of NASA's Astrophysics Data System (ADS) Bibliographic Services and the ArXiv.

\end{acknowledgments}

\bibliographystyle{aasjournal}
\bibliography{allbib}
\end{document}